\documentclass[prl,twocolumn,showpacs,amsmath,amssymb,superscriptaddress]{revtex4-1}
\usepackage{graphicx}
\usepackage{epsfig,epstopdf} 
\usepackage{dcolumn}
\usepackage{bm}
\usepackage{bbm}
\usepackage{ulem}
\usepackage{color}
\usepackage{slashed}
\usepackage{hyperref}
\usepackage{mathrsfs}
\usepackage{subfigure}
\usepackage{verbatim}
\usepackage{dsfont}
\usepackage{float}
\usepackage{amsmath}
\usepackage{tikz-cd}
\usepackage{booktabs}

\newcommand{\nc}{\newcommand}
\nc{\be}{\begin{equation}} \nc{\ee}{\end{equation}}
\nc{\bea}{\begin{eqnarray}} \nc{\eea}{\end{eqnarray}}
\nc{\bean}{\begin{eqnarray*}} \nc{\eean}{\end{eqnarray*}}
\nc{\dg}{\dagger}
\nc{\ua}{\uparrow} \nc{\da}{\downarrow}

\begin{document}

\bibliographystyle{apsrev4-1}

\title{Crystalline symmetry protected Majorana mode in number-conserving Dirac semi-metal nanowires}
\author{Rui-Xing Zhang}
\author{Chao-Xing Liu}
\affiliation{Department of Physics, The Pennsylvania State University, University Park, Pennsylvania 16802}
\date{\today}

\begin{abstract}
One of the cornerstones for topological quantum computations is Majorana zero mode, which has been intensively searched in fractional quantum Hall systems and topological superconductors.
Several recent works suggest that such exotic mode can also exist in one dimensional (1D) interacting double-wire setup even without long-range superconductivity. A notable instability in these proposals comes from inter-channel single-particle tunneling that spoils the topological ground state degeneracy. Here we show that 1D Dirac semimetal (DSM) nanowire is an ideal number-conserving platform to realize such Majorana physics. By inserting magnetic flux, a DSM nanowire is driven into 1D crystalline-symmetry-protected semimetallic phase. Interaction enables the emergence of boundary Majorana zero modes, which is robust as a result of crystalline symmetry protection. We also explore several experimental consequences of Majorana signals.
\end{abstract}


\maketitle

\textit{Introduction -}
Anyons are natural generalizations of bosons and fermions from the perspective of quantum statistics. Interchanging a pair of anyons can induce either a non-trivial phase factor $e^{i\theta}\neq \pm 1$ in the wavefunction (Abelian anyons),
or a rotation operation of the corresponding many-body wave
function among a degenerate set of locally indistinguishable states (non-Abelian anyons) \cite{nayak2008}.
Anyonic physics was first pointed out in the context of fractional quantum Hall
(FQH) effect \cite{tsui1982}, where anyons emerge as bulk quasiparticle excitations in an FQH system. A well-known example here is Majorana quasiparticle (Ising anyon), which emerges in a $\nu=\frac{5}{2}$ Moore-Read FQH state \cite{moore1991}.
The non-Abelian statistics of Majorana quasiparticle makes it a promising candidate for building a topological quantum computer \cite{kitaev2003}. Besides FQH systems, Majorana physics was also studied in the topological superconductor (TSC) after the pioneering works
by Read and Green \cite{read2000}, Ivanov \cite{ivanov2001} and Kitaev \cite{kitaev2001}. In particular, Kitaev pointed out
the existence of boundary Majorana zero mode (MZM) in a one-dimensional (1D) p-wave TSC.
Such TSC is topologically distinct from a conventional superconductor due to the MZM-induced ground state degeneracy (GSD)  \cite{alicea2012}. The degenerate ground states are further labeled by $Z_2$ fermion parity of the system, and their stability is guaranteed by this $Z_2$ parity symmetry. This Kitaev model serves as the underlying mechanism of recent intensive experimental efforts
in realizing MZM physics in 1D semiconductor devices \cite{lutchyn2010,oreg2010,mourik2012,das2012,nadj2014,zhang2017}.

\begin{figure}[t]
  \centering
  \includegraphics[width=0.49\textwidth]{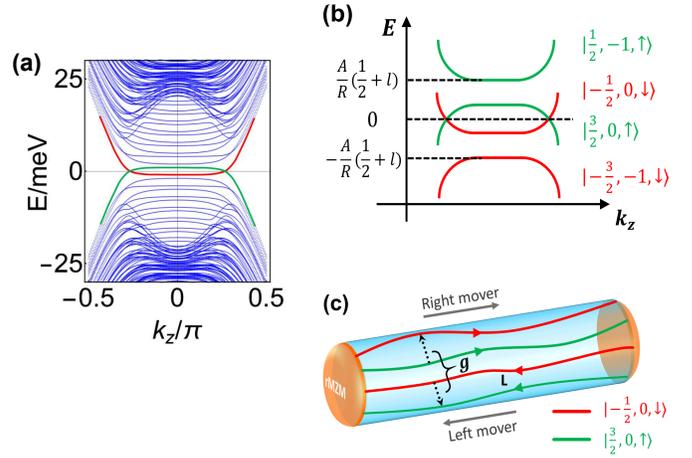}
  \caption{In (a), DSM nanowire is driven into a rotation-symmetry-protected 1D semimetal when flux $\Phi=l\Phi_0$ ($\frac{1}{2}<l<\frac{3}{2}$) is inserted. The emerging 1D Dirac points originate from $|\frac{3}{2},0,\uparrow\rangle$ (green line) and $|-\frac{1}{2},0,\downarrow\rangle$ (red line), as shown in (b). In (c), we show the process of pair-hopping interaction $g$, where two ``red" electrons hop to the ``green" electron states simultaneously. This process respects four-fold rotation symmetry, and enables the emergence of Majorana end states. }
  \label{Fig: DSM nanowire effective}
\end{figure}

Theoretically, it was pointed out that MZM will become unstable in a single 1D quantum wire if strong quantum fluctuations destroy long-range superconductivity \cite{fidkowski2011}. For a double-wire setup, however, MZMs can co-exist with quantum fluctuations
when inter-wire single-particle hopping vanishes
while pair hopping interaction dominates \cite{cheng2011,fidkowski2011,sau2011,kraus2013,ortiz2014,klinovaja2014,iemini2015,lang2015,chen2017,iemini2017}.
Pair hopping process fluctuates particle number of each quantum wire only by a multiple of 2. Thus in each quantum wire, an emergent $Z_2$ fermion parity $P^{(2)}$ is well defined. Consequently, $P^{(2)}$ defines doubly degenerate ground states, which mimics the physics in Kitaev model. However, a well-known issue in the double-wire setup comes from ``$Z_2$ parity breaking" induced by inter-wire single particle tunneling, which is generally unavoidable. Such tunneling process explicitly violates $P^{(2)}$ symmetry and thus spoils Majorana physics.

In this work, we demonstrate how crystalline symmetries naturally solve the ``$P^{(2)}$ breaking" issue, and thus stabilize the Majorana physics without long-range superconductivity. We show that 1D nanowire of three dimensional Dirac semimetals (DSM) \cite{wang2012,wang2013} possesses crystalline-symmetry-protected gapless Dirac points, offering us a material realization of stable semimetallic phase (Fig. \ref{Fig: DSM nanowire effective} (a)). Therefore, a DSM nanowire manifests itself as an effective double-wire setup, with each ``quantum wire" labeled by a representation of the crystalline symmetry group. Inter-``wire" pair-hopping interaction drives the system into an interaction-enabled topological phase with doubly degenerate ground states and MZMs (Fig. \ref{Fig: DSM nanowire effective} (c)). On the other hand, the origin of $P^{(2)}$-breaking, inter-``wire" {\it single}-particle tunneling, is forbidden in our rotation-invariant system, since it explicitly breaks the rotation symmetry. As a result, {\it rotational symmetry protects the robustness of the Majorana physics in the DSM nanowire system.} The boundary MZM in our setup bridges between ground states with different angular momentum representations of the rotational crystalline group.
It is thus dubbed ``representation MZM" (rMZM) to distinguish from conventional MZMs in TSCs. Experimentally, rMZM is shown to exhibit exponentially localized zero-bias peak signal, which can be detected via scanning tunneling microscopy (STM). We also propose a feasible setup to explore the transport physics of rMZM, where the transport phase diagram and experimental signals are discussed.

\textit{1D Dirac points in DSM nanowires -} The following $k\cdot p$ Hamiltonian describes a typical DSM protected by the $C_4$ symmetry\cite{wang2012,wang2013}
\bea
H_{k\cdot p}
=
\begin{pmatrix}
M(k) & Ak_- & 0 & 0 \\
Ak_+ & -M(k) & 0 & 0 \\
0 & 0 & -M(k) & -Ak_- \\
0 & 0 & -Ak_+ & M(k) \\
\end{pmatrix},
\label{Eq: Effective DSM model}
\eea

The basis function $\Psi$ is $(|P,\frac{3}{2}\rangle,|S,\frac{1}{2}\rangle,|S,-\frac{1}{2}\rangle,|P,-\frac{3}{2}\rangle)^T$, where spin-orbit-coupled angular momentum $J\in\{\pm\frac{1}{2},\pm\frac{3}{2}\}$ acts as a pseudo-spin index. Notice that $H_{k\cdot p}$ takes a block-diagonal form of $\text{diag}(H_{\uparrow},H_{\downarrow})$, with each block describing a Weyl Hamiltonian in the corresponding spin ($\uparrow$ or $\downarrow$) sector. Here $M(k)=M_0-M_1 k_z^2-M_2(k_x^2+k_y^2)$ and $k_{\pm}=k_x\pm ik_y$, with $A>0$ and $M_{0,1,2}>0$. The bulk Dirac points are aligned along rotational invariant $k_z$ axis at $k_z=\pm\sqrt{M_0/M_1}$.
The $\uparrow$ and $\downarrow$ sectors are related via $H_{\downarrow}(M(k),A)=H_{\uparrow}(-M(k),-A)$. Thus, we will focus on $H_{\uparrow}$, and the properties of $H_{\downarrow}$ can be obtained analogously.

The DSM nanowire can be better described in cylindrical coordinates. We rewrite $H_{\uparrow}$ in terms of $k_r=-i\partial_r$ and $k_{\theta}=-\frac{i}{r}\partial_{\theta}$ with angular variables $r=\sqrt{x^2+y^2}$ and $\theta=\arctan\frac{y}{x}$. Solving the eigen-state problem \cite{supplementary}, the low-energy eigenstates of $H_{\uparrow}$ are found to be Fermi arc states on the side surface of the nanowire \cite{imura2011}. For a nanowire with radius R, surface Fermi arc spectrum $E_{\uparrow}=-\frac{A}{R}(m+\frac{1}{2})$, where $m=0,\pm1,\pm2,...$ is the eigen-value of $k_{\theta}$. The spin-down part of $H_0$ behaves similarly with $E_{\downarrow}=\frac{A}{R}(m+\frac{1}{2})$. Notice that the complete Fermi arc spectrum always exhibits a finite gap of $A/R$, which originates from the $\pi$ spin Berry phase of the Fermi arc states \footnote{In the limit $R\rightarrow \infty$, the energy gap in both spin sectors approaches zero and DSM nanowire evolves into a bulk DSM sample with two bulk Dirac points connected by Fermi arc states.}. In this nanowire geometry, total angular momentum $J_{tot}$ is a good quantum number, while it is actually composed of two parts: (1) angular contribution from $k_{\theta}$ and (2)
pseudo-spin $J$ which is encoded in the basis $\Psi$. In particular, we find that a state with $k_{\theta}=m$ carries $J_{tot}=m+2\sigma+\frac{1}{2}$ in the spin-$\sigma$ ($\sigma=\pm\frac{1}{2}$) sector. Thus, we label an energy eigen-state as $|J_{tot},m,\sigma\rangle$ where spin index $\sigma\in \{\uparrow,\downarrow\}$.

The 1D Dirac points can be realized by inserting magnetic flux to remove the Berry phase effect. The applied magnetic field should be precisely aligned along the nanowire to preserve the $C_4$ symmetry. With flux $\Phi=l$ (in units of $\Phi_0=h/e$) inserted, $E_{\uparrow/\downarrow}=\mp\frac{A}{R}(m+\frac{1}{2}-l)$. 
The $\pi$ Berry phase is exactly canceled, when $\pi$-flux ($l=\frac{1}{2}$) is inserted. Consequently, $|\frac{3}{2},0,\uparrow\rangle$ touches $|-\frac{1}{2},0,\downarrow\rangle$ at $k_z=0$, which is similar to the worm-hole effect of topological insulator nanowire \cite{rosenberg2010}. When $\Phi$ is further increased, $|\frac{3}{2},0,\uparrow\rangle$ intersects with $|-\frac{1}{2},0,\downarrow\rangle$ to form a gapless inverted band structure (1D Dirac points), as shown in Fig. \ref{Fig: DSM nanowire effective} (b). Since $|\frac{3}{2},0,\uparrow\rangle$ and $|-\frac{1}{2},0,\downarrow\rangle$ belong to different representations of the rotational group, the 1D Dirac points are robust and thus protected by $C_4$ symmetry. In Fig. \ref{Fig: DSM nanowire effective} (a), we verify the above results in the tight-binding model obtained from regularizing the $k\cdot p$ Hamiltonian on a cubic lattice \footnote{The nanowire configuration is modeled by taking periodic (open) boundary condition along the $z$ ($x$ and $y$) direction. Realistic parameters of Cd$_2$As$_3$ are applied in the calculation, and the cross section in the $x$-$y$ plane is chosen to be a $16\times 16$ square}. The magnetic field required for inducing $2\pi$ flux is around 1T for a nanowire with a diameter of 100 nm \cite{supplementary}. This is how we assemble a 1D rotation-symmetry-protected semimetal by inserting magnetic flux into the DSM nanowire.

\textit{Interaction-induced Majorana physics -} The low-energy theory of 1D Dirac points is well captured by the 2-channel Luttinger liquid (LL) theory,
\bea
H_0=\sum_{s=1,2} \int dx \frac{v}{2}[\psi^{\dagger}_{s,R}(x)\partial_x\psi_{s,R}(x)-\psi^{\dagger}_{s,L}(x)\partial_x\psi_{s,L}(x)]. \nonumber
\eea
Here $\psi^{\dagger}_{1,R(L)}$ creates a right (left) moving electron with $J_{tot}=-\frac{1}{2}$, while $\psi^{\dagger}_{2,R(L)}$ creates a right (left) moving electron with $J_{tot}=\frac{3}{2}$.
Two-particle processes that preserve both $U(1)$ charge conservation and $C_4$ rotation symmetry \cite{note2017} are:
\bea
H_1&=&\int dx [g\psi^{\dagger}_{1,R}\psi^{\dagger}_{1,L}\psi_{2,R}\psi_{2,L}
+g_1\psi^{\dagger}_{1,R}\psi^{\dagger}_{2,L}\psi_{2,R}\psi_{1,L} \nonumber \\
&&+g_2\psi^{\dagger}_{1,R}\psi^{\dagger}_{2,R}\psi_{2,L}\psi_{1,L}+h.c.]
\label{Eq: C4 interaction}
\eea
If we move the Fermi level slightly away from the Dirac points (zero energy in Fig. \ref{Fig: DSM nanowire effective} (a) and (b)), both $g_1$ and $g_2$ scatterings involve certain amount of momentum transfer, and will be suppressed in a translational invariant system. Inter-channel pair-hopping $g$, however, preserves both momentum conservation and $C_4$ symmetry (angular momentum transfer $\Delta J_{tot}=4$). As will be shown below, it is $g$ that is responsible for Majorana physics.

We next apply Abelian bosonization technique \cite{giamarchi2004} and define $\psi_{s,R}\sim e^{i\sqrt{\pi}(\phi_s-\theta_s)}$ and $\psi_{s,L}\sim e^{-i\sqrt{\pi}(\phi_s+\theta_s)}$ following the convention in \cite{fradkin2013}. It is convenient to introduce the bonding and anti-bonding fields as $\phi_{\pm}=\frac{1}{\sqrt{2}}(\phi_1\pm \phi_2)$ and $\theta_{\pm}=\frac{1}{\sqrt{2}}(\theta_1\pm \theta_2)$, as well as $K_{\pm}$ to be the Luttinger parameters for bonding and anti-bonding channel. The bonding sector remains gapless, while the anti-bonding sector might open up a non-trivial gap with
the Hamiltonian
\bea
H_-&=&\int dx \frac{v}{2}[K_-(\partial_x\phi_-)^2+\frac{1}{K_-}(\partial_x\theta_-)^2] \nonumber \\
&&+g\int dx \cos2\sqrt{2\pi}\theta_-.
\label{Eq: H_-}
\eea
We will focus on $K_-<1$ where $g$ is relevant under renormalization group analysis. At Luther-Emery point ($K_-=\frac{1}{2}$), Eq. \ref{Eq: H_-} can be exactly mapped to the topological Kitaev model with MZM end states \cite{cheng2011,supplementary}. Such mapping is achieved by refermionizing Eq. \ref{Eq: H_-} with $\tilde{\psi}_R\sim e^{i\sqrt{\pi}(\tilde{\phi}-\tilde{\theta})}$ and $\tilde{\psi}_L\sim e^{-i\sqrt{\pi}(\tilde{\phi}+\tilde{\theta})}$, where $\tilde{\phi}=\phi_-/\sqrt{2}$ and $\tilde{\theta}=\sqrt{2}\theta_-$. Away from $K_-=\frac{1}{2}$, the above mapping fails while we will show that Majorana physics (both GSD and Majorana end states) persists.

The ground state is obtained by minimizing $\cos2\sqrt{2\pi}\theta_-$ and pinning $\theta_-=(n_{\theta}+\frac{1}{2})\sqrt{\pi/2}$, where $n_{\theta}\in \mathbb{Z}$ is an integer-valued operator. Since $\theta_-$ has $\sqrt{2\pi}$ periodicity, there are two degenerate ground states $|\theta_-=\pm\frac{1}{2}\sqrt{\frac{\pi}{2}}\rangle$, which are characterized by the emergent $Z_2$ fermion parity $P^{(2)}$. Physically, $P^{(2)}$ counts the parity of electron number in the $J_{tot}=-\frac{1}{2}$ subspace,
\bea
P^{(2)}&=&(-1)^{\sqrt{\frac{1}{\pi}}\int_0^L dx \partial_x\phi_1}=P_+P_-,
\eea
where a nanowire with finite length L is considered and $P_{\pm}=e^{i\sqrt{\pi/2}(\phi_{\pm}(L)-\phi_{\pm}(0))}$. With $P^{(2)}\theta_- (P^{(2)})^{-1}=\theta_--\sqrt{\frac{\pi}{2}}$, $P^{(2)}$ interchanges $|\theta_-=\pm\frac{1}{2}\sqrt{\frac{\pi}{2}}\rangle$ from one to another. Quantum superposition principle, however, allows us to define the following degenerate ground states,
\bea
|\pm\rangle&=&\frac{1}{\sqrt{2}}[|\theta_-=\frac{1}{2}\sqrt{\frac{\pi}{2}}\rangle\pm|\theta_-=-\frac{1}{2}\sqrt{\frac{\pi}{2}}\rangle],
\eea
where $|+\rangle$ and $|-\rangle$ are characterized by even and odd $P^{(2)}$ parity, respectively. Since $P^{(2)}$ is a global property of the system, the degeneracy here is topological, which can NOT be distinguished via any local measurement.

Topological GSD can also be revealed by constructing rMZM operators explicitly. Imposing open boundary condition at $x=0$ gives rise to $\psi_{l,L}(0)+\psi_{l,R}(0)=0$ for $l=1,2$,
which corresponds to $\phi_l(0)=n_l^{(1)}\sqrt{\pi}$ up to an unimportant constant. Here, $n_l^{(1)}\in \mathbb{Z}$ is an integer-valued operator. Introducing $n_+^{(1)}=n_1^{(1)}+n_2^{(1)}$ and $n_-^{(1)}=n_1^{(1)}$, the boundary condition fixes the value of $\phi_{\pm}$ as $\phi_+(0)=n_+^{(1)}\sqrt{\pi/2}$ and $\phi_-(0)=(2n_-^{(1)}-n_+^{(1)})\sqrt{\pi/2}$. It is important to notice that $[n_-^{(1)}(x),n_{\theta}(x')]=\frac{i}{\pi}\Theta(x-x')$, while $n_+^{(1)}$ always commutes with $n_{\theta}$ and behaves like a c-number. Following Ref. \cite{clarke2013}, we construct rMZM operator $\alpha_1$ at $x=0$ and $\alpha_2$ at $x=L$ as,
\bea
\alpha_1=e^{i\pi (n_-^{(1)}+n_{\theta})},\\ \alpha_2=e^{i\pi (n_-^{(2)}+n_{\theta})},
\eea
where we have defined the boundary condition at $x=L$ to be $\phi_+(L)=n_+^{(2)}\sqrt{\pi/2}$ and $\phi_-(L)=(2n_-^{(2)}-n_+^{(2)})\sqrt{\pi/2}$ in a similar way. The Majorana properties of $\alpha_{1,2}$ can be easily checked, where $[\alpha_{1,2},H_-]=0$ and $\alpha_1^2=\alpha_2^2=1$. Starting from a ground state $|+\rangle$, one can easily show that $\alpha_{1,2}|+\rangle=|-\rangle$ up to some phase factors. This also proves the topological GSD.

The essential role of $C_4$ symmetry in protecting rMZM should be emphasized. $C_4$ symmetry enforces inter-channel single-particle tunneling events to vanish as they break $C_4$ explicitly, which makes $P^{(2)}$ symmetry well-defined. Since $C_4\theta_- C_4^{-1}=\theta_-+\sqrt{\frac{\pi}{2}}$, it is easy to check that $C_4|\pm\rangle=\pm|\pm \rangle$, which proves $C_4=P^{(2)}$. The equivalence between a $N$-fold discrete symmetry $Z_N$ and an emergent $Z_M$ fermion parity $P^{(M)}$ is a quite general result for two-channel systems, which is discussed in details in the supplementary materials \cite{supplementary}.

We stress that our theory is not limited to the $C_4$-symmetric DSM nanowire. For a 1D nanowire growing along the $x$ direction, the crystalline structure in the cross-section ($y$-$z$ plane) is characterized by a two-dimensional (2D) point group $G_{2D}$, where every symmetry operation in $G_{2D}$ leaves the $x$ direction invariant. Therefore, it is natural to generalize our theory from $C_4$ group to general 2D point groups. In \cite{supplementary}, we identify simple criteria to determine whether a given symmetry group $G$ could give rise to $P^{(2)}$ parity and present a complete discussion of all 2D point groups, as well as their double groups. In particular, our theory predicts all possible irreducible representations for each $G_{2D}$ that could support Majorana physics, and thus establishes the guideline for the search of other candidate materials.

\begin{figure*}[t]
  \centering
  \includegraphics[width=0.7\textwidth]{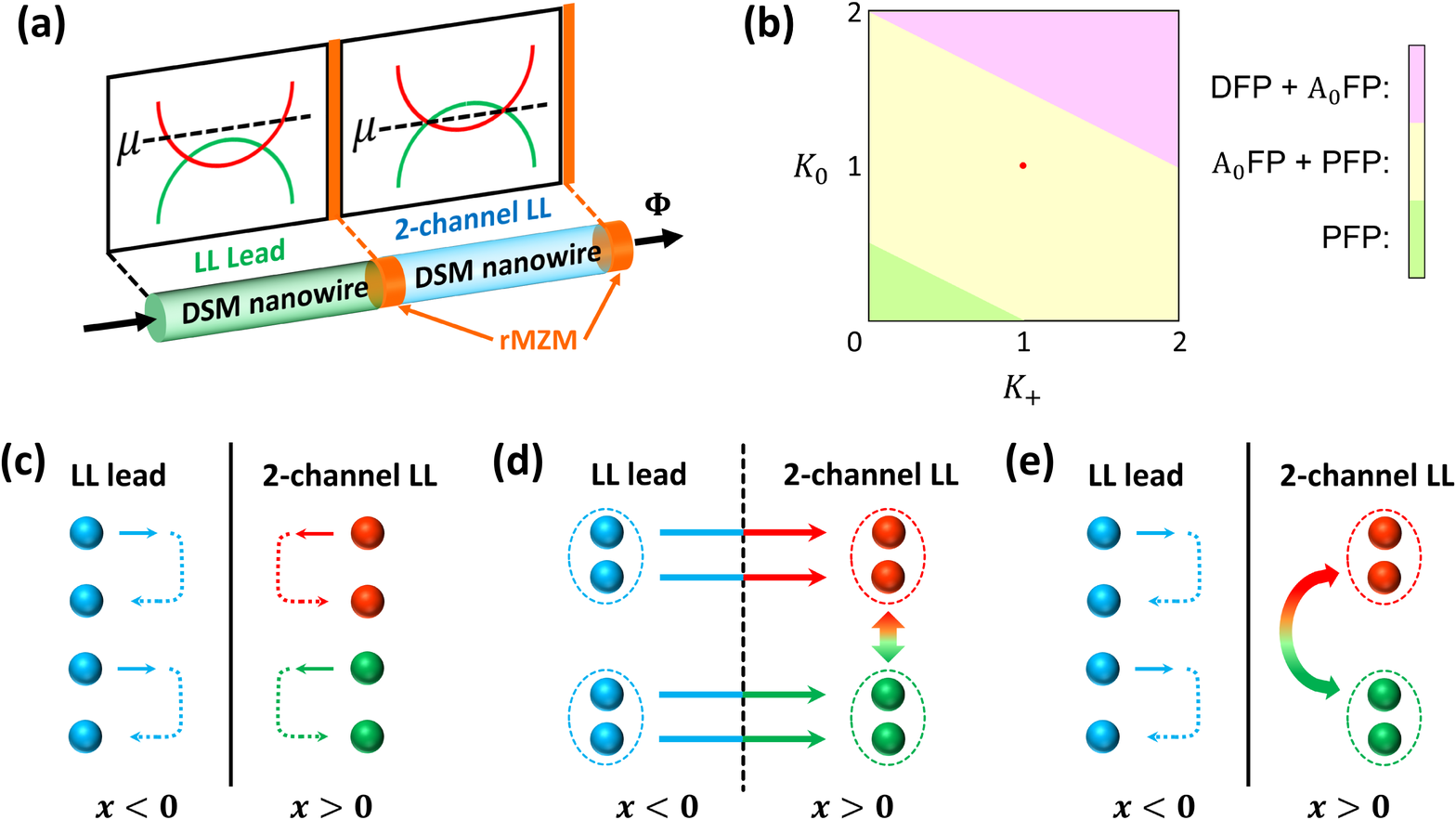}
  \caption{(a) We apply a step-function gate to DSM nanowire to fabricate the 2LL/LL configuration. (b) Tunneling phase diagram of 2LL/LL junction with $K_-<1$. In (c) - (e), we systematically plot the boundary conditions of the following fixed points: DFP, PFP and A$_0$FP. The dashed circle that contains a pair of electrons signals pair-tunneling/backscattering process. }
  \label{Fig:transport}
\end{figure*}

\textit{Experimental detection -} One important approach to probe Majorana signals is to map out the spatial profile of spectral density using STM technique, and seek for exponentially localized zero-bias peak that originates from MZM bound state. Following this logic, we consider the single-electron-tunneling problem from a Fermi liquid lead to the DSM nanowire and seek for rMZM signals. This tunneling process changes both electron number and $P^{(2)}$ simultaneously. In particular, the ground state is changed from $|0\rangle=|N,p\rangle$ with $N$ electrons and $P^{(2)}=p$ to $|1\rangle=|N+1,-p\rangle$ with $N+1$ electrons and $P^{(2)}=-p$ via this tunneling process. For a DSM nanowire at $x\in[0,L]$, in the $L\rightarrow \infty$ limit, transition matrix element of injecting a single electron at $x=x_0$ ($0\leq x_0\ll L$) can be calculated with the help of mode expansion technique \cite{keselman2015,supplementary}, and the resulting spectral density is
\bea
\rho(x,\omega\rightarrow 0)&=&|\langle 1|\psi_{1,R}^{\dagger}|0\rangle|^2
= {\cal N}(x,\epsilon,K_{\pm})\ e^{-\frac{\pi}{4K_-}\frac{x}{\xi}}
\label{Eq: Spectral density}
\eea
where $\xi=\sqrt{\frac{v}{8\pi g K_-}}$ is the correlation length of the system and $\epsilon$ is the short distance cut-off. While ${\cal N}(x,\epsilon,K_{\pm})$ counts the power-law contribution of this transition process \cite{supplementary}, the exponential part of $\rho(x,\omega\rightarrow 0)$ unambiguously reveals an exponentially localized rMZM on the boundary, which is ready to be detected using STM technique. Since $\xi\sim 1/\sqrt{gK_-}$, the stronger the interaction, the more localized rMZM will be.

Finally, we consider a 2-channel LL/LL lead (2LL/LL) junction, which can be experimentally achieved by applying a step-function-like gating to the DSM nanowire, as shown in Fig. \ref{Fig:transport} (a). With this gating configuration, the right part of the nanowire ($x>0$) is in the 2-channel LL regime, while the left part ($x<0$) becomes a single-channel LL that plays the role of a lead. The rMZM is expected at the interface between 2-channel LL and single-channel LL ($x=0$).

The transport physics of 2LL/LL junction is characterized by tunneling fixed points. Fig. \ref{Fig:transport} (c) - (e) depict three stable fixed points for different physical boundary conditions at the interface. (i) Disconnecting fixed point (DFP) signals the promotion of normal reflection process, where electrons will be perfectly reflected at the interface between 2LL and LL (Fig. \ref{Fig:transport} (c)). (ii) Pair-tunneling fixed point (PFP) characterizes the resonant Andreev reflection process \cite{law2009}, where electrons will pair up and tunnel through the interface without any backscattering, as shown in Fig. \ref{Fig:transport} (d). (iii) $A_0$ fixed point ($A_0$FP) also signals normal reflection process of electrons from the LL lead, while on the 2LL side, electrons of channel-$1$ (channel-$2$) will pair up and be scattered together into channel-$2$ (channel-$1$) at the interface, as shown in Fig. \ref{Fig:transport} (e). This is essentially different from the case of DFP, where electrons will be individually backscattered without any channel-switching. We notice that DFP and PFP have been previously discussed in a TSC/LL junction \cite{fidkowski2012}, while $A_0$FP is a new fixed point in our system.

With the ``delayed evaluation of boundary condition" (DEBC) method \cite{oshikawa2006,hou2012}, we verified the existence of all three fixed points (DFP, PFP, and A$_0$FP) and evaluate the scaling dimensions of perturbation terms at each fixed point. A brief introduction of this method and details of fixed point information can be found in the supplementary materials \cite{supplementary}. The complete phase diagram is mapped out in Fig. \ref{Fig:transport} (b) for $K_-<1$. In the strongly attractive regime ($K_+<1-2K_0$), PFP is stable and promotes the Andreev reflection process. In the strongly repulsive limit ($K_+>4-2K_0$), both DFP and A$_0$FP are stable, and the tunneling conductance is suppressed.

A realistic system is most likely to fall into the weakly interacting regime ($K_0,K_+\approx 1$) with both stable PFP and $A_0$FP. In a transport measurement, one expects a zero-bias conductance peak at PFP, while a vanishing conductance at $A_0$FP. The co-existence of two stable fixed points suggests the emergence of an intermediate unstable fixed point which characterizes the transition between PFP and $A_0$FP. This novel phase transition is intriguing, which is definitely worthy to be explored in future works. At last, the role of rMZM is explored by mapping out a Majorana-free phase diagram where we quench pair-hopping interaction $g$ at $x>0$. In this case, we find a completely different phase diagram with more exotic phases \cite{supplementary}. Especially, the system falls into $A_0$FP near $K_{0,+}\approx 1$, while PFP only shows up in the strongly attractive regime. Therefore, the appearance of weakly interacting PFP in Fig. \ref{Fig:transport} (b) is a direct consequence of rMZM, and serves as a transport evidence of rMZM.

\textit{Discussion -} In summary, we have proposed that magnetic-flux insertion drives DSM nanowire into a 1D crystalline-symmetry-protected semimetal, which serves as an ideal platform to realize Majorana physics without long-range superconductivity. In particular, crystalline symmetry forbids inter-channel single-particle tunneling, and thus guarantees the stability of Majorana physics. We notice that DSM nanowire of Cd$_3$As$_2$ has been successfully fabricated \cite{li2015,wang2016}, while these nanowires have been grown along $[112]$ direction, so that $C_4$ rotation symmetry is explicitly broken and fails to support rMZM. Other promising candidate materials include heterostructures of Kondo materials \cite{ok2017}, where a correlated DSM phase protected by $C_4$ symmetry is found.
Finally, we emphasize once again that our theory is general and not limited to the DSM nanowires. The classification of 2D point groups in the supplementary materials will inspire more future efforts into realizing symmetry-protected Majoranas in number conserving systems.

\textit{Acknowledgement -} The authors are indebted to Meng Cheng for valuable suggestions. R.-X.Z would like to thank Jian-Xiao Zhang, Jiabin Yu and Jiahua Gu for helpful discussions, and particularly Lun-Hui Hu for collaboration on a closely related project.
C.-X. L. and R.-X.Z acknowledge support from Office of Naval Research (Grant No. N00014-15-1-2675).

\bibliography{DSM}

\begin{thebibliography}{53}
\bibitem [{not()}]{note2017}%
  \BibitemOpen
  \href@noop {} {\bibinfo  {journal} {Notice that the effective $k\cdot p$
  Hamiltonian in Eq. \ref{Eq: Effective DSM model} has the full-rotation $O(2)$
  symmetry, since we focused on the continuum limit and dropped higher order
  terms. However, when we consider interaction effects, it is important to
  distinguish this artificial $O(2)$ symmetry from the realistic $C_4$
  symmetry. This is because O(2) is a stronger symmetry and also imposes a
  stronger constraint to interaction terms. In particular, the pair-hopping
  term $g$ that leads to Majorana physics transfers angular momentum by $\Delta
  J_{tot}=4$. As a result, pair-hopping interaction only preserves $C_4$
  symmetry, but explicitly breaks O(2) symmetry}\ }\BibitemShut {NoStop}%
\end{thebibliography}

\onecolumngrid

\subsection{\large Supplementary Materials for ``Crystalline symmetry protected Majorana mode in number-conserving Dirac semi-metal nanowires"}

\section{Appendix A. Dirac semimetal in cylindrical coordinates}
In this appendix, we will follow Ref. \cite{imura2011} to derive the low-energy theory of DSM nanowire in a cylindrical geometry. A typical Hamiltonian for Dirac semimetals (such as Na$_3$Bi and Cd$_3$As$_2$) is given by \cite{wang2012,wang2013}
\bea
H_{Dirac}=\begin{pmatrix}
	H_{\uparrow} & 0 \\
	0 & H_{\downarrow} \\
\end{pmatrix}
=
\begin{pmatrix}
	M(k) & Ak_- & 0 & 0 \\
	Ak_+ & -M(k) & 0 & 0 \\
	0 & 0 & -M(k) & -Ak_- \\
	0 & 0 & -Ak_+ & M(k) \\
\end{pmatrix}.
\label{Eq: Effective DSM model}
\eea
The basis functions are $|P,\frac{3}{2}\rangle,|S,\frac{1}{2}\rangle,|S,-\frac{1}{2}\rangle,|P,-\frac{3}{2}\rangle$. Here $M(k)=M_0-M_1 k_z^2-M_2(k_x^2+k_y^2)$ and $k_{\pm}=k_x\pm ik_y$, with $A>0$ and $M_{0,1,2}>0$. The bulk Dirac points are $k_z=\pm K_0=\pm\sqrt{\frac{M_0}{M_1}}$. Notice that $H_{\downarrow}(M(k),A)=H_{\uparrow}(-M(k),-A)$ and we will focus on $H_{\uparrow}$ in the following discussion. In cylindrical coordinate, $(x,y,z)\rightarrow (r,\theta,z)$ with
\bea
r=\sqrt{x^2+y^2},\ \theta=\arctan\frac{y}{x}.
\eea
It is easy to show that
\bea
k_+&=&e^{i\theta}(k_r+ik_{\theta}) \nonumber \\
k_-&=&e^{-i\theta}(k_r-ik_{\theta})
\eea
with $k_r=-i\frac{\partial}{\partial r}$ and $k_{\theta}=-i\frac{1}{r}\frac{\partial}{\partial \theta}$. When the radius $r$ is much larger than the lattice constant $a$ with $r\gg a$, we also have $k_x^2+k_y^2=k_r^2+k_{\theta}^2$. Then $H_{\uparrow}$ can be separated into $H_{\uparrow,\perp}$ describing the physics normal to the surface and $H_{\uparrow,\parallel}$ describing the physics parallel to the surface. Let us consider the physics around some $k_0$ on the $k_z$ axis, with $k_z=k_0+\delta k_z$. For simplicity, we will use $k_z$ to represent $\delta k_z$ in the later discussion. We find that
\bea
H_{\uparrow,\perp}=
\begin{pmatrix}
	N(k_0)-M_2k_r^2 & Ae^{-i\theta}k_r \\
	Ae^{i\theta}k_r & -N(k_0)+M_2k_r^2 \\
\end{pmatrix}
\eea
and
\bea
H_{\uparrow,\parallel}=
\begin{pmatrix}
	-2M_1k_0k_z-M_1k_z^2-M_2k_{\theta}^2 & -iAe^{-i\theta}k_{\theta} \\
	iAe^{i\theta}k_{\theta} & 2M_1k_0k_z+M_1k_z^2+M_2k_{\theta}^2 \\
\end{pmatrix}.
\eea
Here $N(k_0)=M_0-M_1k_0^2$. The strategy here is to solve for eigenstate of $H_{\uparrow,\perp}$ with open boundary condition and use this result to project $H_{\uparrow,\parallel}$ to extract low-energy effective theory of the Dirac semimetal nanowire. Let us assume the radius of the nanowire is $R\gg a$, and the open boundary condition is $\psi_{\perp}(r=R)=0$. Then a trial wavefunction of $H_{\uparrow,\perp}$ should take the form
\bea
\psi_{\perp,\lambda}(r)\sim e^{\lambda(r-R)}
\begin{pmatrix}
	C \\ De^{i\theta}\\
\end{pmatrix}
\eea
Then the energy eigenstate with energy $E$ can be shown to take the form
\bea
\psi_{\perp,\lambda}(E,r)\sim e^{\lambda(r-R)}
\begin{pmatrix}
	E+m(\lambda) \\ -i\lambda A e^{i\theta}\\
\end{pmatrix}
\eea
where $m(\lambda)=N_0+M_2\lambda^2$. On the other hand, eigen-equation now becomes
\bea
\det \begin{pmatrix}
	m(\lambda)-E & -iAe^{-i\theta}\lambda \\
	-iAe^{i\theta}\lambda & -m(\lambda)-E
\end{pmatrix}=0
\eea
$\lambda_{\pm}$ are the two positive $\lambda$ solutions of the above equation, and they satisfy
\bea
\lambda_+\lambda_-=\sqrt{\frac{N_0^2-E^2}{M_2^2}}
\eea
Notice that these solutions only exist when $|E|<|N_0|$. This implies that the surface state solutions are in-gap states which live at $-k_0<k_z<k_0$. The wavefunction now can be written as a linear superposition of $\lambda_+$ and $\lambda_-$ eigenstates,
\bea
\psi_{\perp}(E,r)=c_+e^{\lambda_+(r-R)}\psi_{\perp,\lambda_+}+c_-e^{\lambda_-(r-R)}\psi_{\perp,\lambda_-}
\eea
The boundary condition at $r=R$ indicates that
\bea
\det\begin{pmatrix}
	m(\lambda_+)+E & m(\lambda_-)+E \\
	-iAe^{i\theta}\lambda_+ & -iAe^{i\theta}\lambda_-
\end{pmatrix}=0
\eea
After some calculations, we find that
\bea
E(E+N_0)=0
\eea
For $|E|<|N_0|$, the only solution is
\bea
E&=&0 \nonumber \\
m(\lambda)&=&\pm A\lambda
\eea
Together with the definition of $m(\lambda)$, $\lambda_{\pm}$ must satisfy
\bea
M_2\lambda^2\mp A\lambda+N_0=0
\eea
or equivalently
\bea
\lambda_+ + \lambda_-&=&\pm\frac{A}{2M_2} \nonumber \\
\lambda_+\lambda_-&=&\frac{N_0}{M_2}
\eea
Since $\lambda_{\pm}>0$, we have $m(\lambda)=A\lambda$ and $N_0>0$, which implies $|k_0|<\frac{M_0}{M_1}=K_0$. Therefore, this in-gap surface states only exists between two bulk Dirac points, and are identified as the Fermi arc states connecting the bulk Dirac points. The surface state wavefunction now takes the form
\bea
\psi_{\perp}(E,r,\theta)=f(\lambda_{\pm},r)
\begin{pmatrix}
	1 \\ -ie^{i\theta}
\end{pmatrix}
\label{Eq: eigenstate}
\eea
Now let us obtain the projected low energy theory of $H_{\uparrow, \parallel}$, and we find that
\bea
H_{\uparrow,eff}&=&\psi_{\perp}^{\dagger}H_{\uparrow,\parallel}\psi_{\perp} \nonumber \\
&=&-\frac{A}{r}(-i\frac{\partial}{\partial \theta}+\frac{1}{2})+\frac{M_2}{r^2}(1-2i\frac{\partial}{\partial\theta})
\eea
With periodic boundary condition along $\hat{\theta}$ direction, $k_{\theta}$ is a good quantum number and $-i\partial_{\theta}$ operator  takes integer values $m=0,1,2,3...$. It is easy to see that the lowest energy state locates at $r=R$, and when $R\gg a$, the second term in $H_{\uparrow,eff}$ can be ignored. We find that
\bea
E_{\uparrow,eff}=-\frac{A}{R}(m+\frac{1}{2}).
\eea
As for the spin-down part, since $H_{\downarrow}(M(k),A)=H_{\uparrow}(-M(k),-A)$, it is easy to see that
\bea
E_{\downarrow,eff}=\frac{A}{R}(m+\frac{1}{2}).
\eea
It is easy to see that the energy spectrum is gapped because of the $\pi$ Berry phase which gives rise to a $\frac{A}{R}$ energy gap.

To remove the Berry phase effect, we apply a magnetic field, which must be carefully aligned along the nanowire to preserve the rotation symmetry of the system. Assume the flux is $\Psi=l\Psi_0$ with $\Psi_0=\frac{h}{e}$ being the flux quantum, the energy spectrum is modified to
\bea
E_{\uparrow,eff}&=&-\frac{A}{R}(m+\frac{1}{2}-l) \nonumber \\
E_{\downarrow,eff}&=&\frac{A}{R}(m+\frac{1}{2}-l)
\eea
Therefore, inserting magnetic flux shifts $E_{\uparrow,eff}$ by $\frac{lA}{R}$ and $E_{\downarrow,eff}$ by $-\frac{lA}{R}$. When $l>\frac{1}{2}$, we find that $|m=0,\uparrow\rangle$ ($m=0$ state in the spin up sector) and $|m=0,\downarrow\rangle$ form gapless inverted band structure, leading to semimetallic dispersions.

As a result of Eq. \ref{Eq: eigenstate}, a state labeled with $m$ can be written as
\bea
|m,\uparrow\rangle&\sim& e^{im\theta}|\psi_{\perp},\uparrow\rangle =e^{im\theta}(|\frac{3}{2}\rangle-ie^{i\theta}|\frac{1}{2}\rangle) \nonumber \\
|m,\downarrow\rangle&\sim& e^{im\theta}|\psi_{\perp},\downarrow\rangle =e^{im\theta}(|-\frac{1}{2}\rangle-ie^{i\theta}|-\frac{3}{2}\rangle)
\eea
where we have ignored its $r$ and $k_z$ dependence. Therefore, even with the same $m$ value, the spin-up state and the spin-down state have different angular momentum. Their angular momentum can be read out when we apply a rotation operator $C_{\theta}=e^{iJ\theta}$, and we find that
\bea
C_{\theta}|m,\uparrow\rangle&=&e^{i(m+\frac{3}{2})\theta}|m,\uparrow\rangle\ \ \text{with }J=m+\frac{3}{2} \nonumber \\
C_{\theta}|m,\downarrow\rangle&=&e^{i(m-\frac{1}{2})\theta}|m,\downarrow\rangle \ \ \text{with }J=m-\frac{1}{2}
\eea
Therefore, $|m=0,\uparrow\rangle$ carries angular momentum $J=\frac{3}{2}$ while $|m=0,\downarrow\rangle$ carries angular momentum $J=-\frac{1}{2}$. These two states belong to different representations of rotation group, and the crossing between them is protected.

In our discussion, we create only one pair of Dirac points with flux number $\frac{1}{2}<l<\frac{3}{2}$. In general, one can keep increasing flux to create more 1D Dirac points, realizing an N-channel LL ($N>2$) system. Such N-channel LL could be a potential realization of coupled wire system, whose interaction effects have been widely discussed \cite{kane2002,teo2014,neupert2014,sagi2014,klinovaja2014a}.

\begin{figure}[t]
	\centering
	\includegraphics[width=0.4\textwidth]{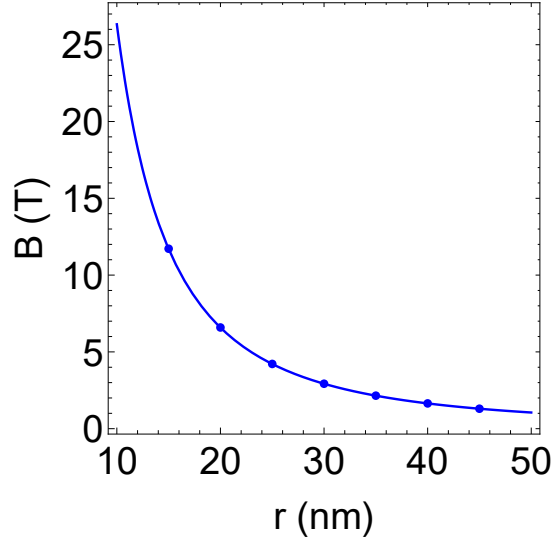}
	\caption{The magnetic field for inducing $2\pi$ flux in a nanowire with a radius $r$.}
	\label{Fig: Magnetic field}
\end{figure}

In Fig. \ref{Fig: Magnetic field}, we numerically calculate the magnetic field for inducing $2\pi$ magnetic flux in a nanowire with a radius $r$. For the existing nanowire setup of Cd$_3$As$_2$ \cite{wang2016} with a diameter of around $100$ nm, the magnetic field for $2\pi$-flux is around $0.8$T.

\section{Appendix B. Microscopic origin of pair-hopping interaction}

In this section, we will discuss the microscopic origin of pair-hopping interaction $g$. Let us recall that the low-energy physics originates from the following channels:
\bea
\text{Channel 1: } |m=0,\downarrow\rangle&\sim& |-\frac{1}{2}\rangle-ie^{i\theta}|-\frac{3}{2}\rangle \nonumber \\
\text{Channel 2: } |m=0,\uparrow\rangle&\sim& |\frac{3}{2}\rangle-ie^{i\theta}|\frac{1}{2}\rangle
\eea
We can expand the angular momentum bases with the bases of spin and atomic orbitals as follows,
\bea
|\frac{3}{2}\rangle&=&\beta_1 |p_+,\uparrow\rangle + ... \nonumber \\
|\frac{1}{2}\rangle&=&\alpha_2 |s,\uparrow\rangle+\beta_2 |p_+,\downarrow\rangle + ... \nonumber \\
|-\frac{1}{2}\rangle&=&\alpha_3|s,\downarrow\rangle + \beta_3 |p_-,\uparrow\rangle +... \nonumber \\
|-\frac{3}{2}\rangle&=&\beta_4|p_-,\downarrow\rangle+...
\eea
where $\alpha_i$ and $\beta_i$ are coefficients that depend on material details. Now, we will start from the pair-hopping interaction $g$ and trace back its microscopic origin in the atomic orbital bases:
\bea
H_1&=&g\int dx \int_0^{2\pi}Rd\theta\  \psi^{\dagger}_{2,R}\psi^{\dagger}_{2L}\psi_{1,R}\psi_{1,L} \nonumber \\
&=&g\int dx \int_0^{2\pi}Rd\theta\ (c^{\dagger}_{\frac{3}{2},R}+ie^{-i\theta}c^{\dagger}_{\frac{1}{2},R})(c^{\dagger}_{\frac{3}{2},L}+ie^{-i\theta}c^{\dagger}_{\frac{1}{2},L})(c_{-\frac{1}{2},R}-ie^{i\theta}c^{\dagger}_{-\frac{3}{2},R})(c_{-\frac{1}{2},L}-ie^{i\theta}c^{\dagger}_{-\frac{3}{2},L}) \nonumber \\
&=& g \int_0^{2\pi}Rd\theta[(V_{3311}+V_{1133}+V_{3113}+V_{1331}+V_{3131}+V_{1313})+e^{-i\theta}V_{-1}+e^{i\theta}V_{1}+e^{-2i\theta}V_{-2}+e^{2i\theta}V_{2}]
\eea
where we have defined
\bea
V_{ijkl}=\int dx\  \psi^{\dagger}_{\frac{i}{2},R}\psi^{\dagger}_{\frac{j}{2},L}\psi_{-\frac{k}{2},R}\psi_{-\frac{l}{2},L}
\eea
The explicit form of $V_{\pm 1(2)}$ are not important to our analysis, since these terms will vanish under the integral of angular variable $\theta$:
\bea
\int_0^{2\pi}e^{in\theta}d\theta=2\pi\delta_{n,0},\ \ \forall\ n\in \mathbb{Z}
\eea
Therefore,
\bea
H_1=2\pi g\int dx (V_{3311}+V_{1133}+V_{3113}+V_{1331}+V_{3131}+V_{1313})
\eea
Let us take $V_{3113}$ as an example. In the momentum space representation,
\bea
V_{3113}&=&\sum_{k,k',q}\beta_1^*\beta_4 c^{\dagger}_{p_+,\uparrow,k}(\alpha_2^* c^{\dagger}_{s,\uparrow,k'}+\beta_2^*c^{\dagger}_{p_+,\downarrow,k'})
(\alpha_3 c_{s,\downarrow,k'+q}+\beta_3 c_{p_-,\uparrow,k'+q}) c_{p_-,\downarrow,k-q} \nonumber \\
&=&\beta_1^*\alpha_2^*\alpha_3\beta_4\sum_q(\sum_k c^{\dagger}_{p_+,\uparrow,k}c_{p_-,\downarrow,k-q})(\sum_{k'}c^{\dagger}_{s,\uparrow,k'}c_{s,\downarrow,k'+q}) \nonumber \\
&&+\beta_1^*\beta_2^*\beta_3\beta_4\sum_q(\sum_k c^{\dagger}_{p_+,\uparrow,k}c_{p_-,\downarrow,k-q})(\sum_{k'}c^{\dagger}_{p_+,\downarrow,k'}c_{p_-,\uparrow,k'+q})+... \nonumber \\
&=&\beta_1^*\alpha_2^*\alpha_3\beta_4\sum_q S^+_{p,\frac{3}{2}} (q)S^+_{s,\frac{1}{2}}(-q)+\beta_1^*\beta_2^*\beta_3\beta_4\sum_q S^+_{p,\frac{3}{2}} (q)S^-_{p,\frac{1}{2}}(-q)+...
\eea
Here we have defined the spin operators as
\bea
S^i_{p,\frac{3}{2}} (q)&=&\sum_k(c^{\dagger}_{p_+,\uparrow,k+q},c^{\dagger}_{p_-,\downarrow,k+q})\tau^i
\begin{pmatrix}
	c_{p_+,\uparrow,k} \\
	c_{p_-,\downarrow,k}
\end{pmatrix} \nonumber \\
S^i_{p,\frac{1}{2}} (q)&=&\sum_k(c^{\dagger}_{p_-,\uparrow,k+q},c^{\dagger}_{p_+,\downarrow,k+q})\tau^i
\begin{pmatrix}
	c_{p_-,\uparrow,k} \\
	c_{p_+,\downarrow,k}
\end{pmatrix} \nonumber \\
S^i_{s,\frac{1}{2}} (q)&=&\sum_k(c^{\dagger}_{s,\uparrow,k+q},c^{\dagger}_{s,\downarrow,k+q})\tau^i
\begin{pmatrix}
	c_{s,\uparrow,k} \\
	c_{s,\downarrow,k}
\end{pmatrix}
\eea
where $\tau^{i=x,y,z}$ is the Pauli matrix. The spin ladder operators are defined as
\bea
S^{\pm}_{\alpha,J}=S^x_{\alpha,J}\pm iS^y_{\alpha,J}
\eea
Physically, the first term in $V_{3113}$ flips two spin simultaneously, which could originate from the combinational effects of Coulomb interaction and spin-orbit coupling. The second term is the conventional multi-orbital Coulomb exchange interaction.

\section{Appendix C. Expressions of $K_+$ and $K_-$}

In this section, we consider the renormalization effects of the following density-density interactions to the Luttinger parameters $K_+$ and $K_-$.
\bea
V_{density}=g_3\sum_{s=1,2}\rho_{s,L}\rho_{s,R}+g_4\sum_{s\neq s'\in\{1,2\}}\rho_{s,L}\rho_{s',R}+g_5 \sum_{s=1,2}(\rho_{s,L}^2+\rho_{s,R}^2)
\eea
Here, $g_3$ and $g_4$ denote intra-channel and inter-channel interactions, respectively. In the bosonization language, density operators are defined as
\bea
\rho_{s,L}&=&\frac{1}{2\sqrt{\pi}}\partial_x(\phi_s+\theta_s) \nonumber \\
\rho_{s,R}&=&\frac{1}{2\sqrt{\pi}}\partial_x(\phi_s-\theta_s)
\eea
Each term of $V_{density}$ can be bosonized into
\bea
g_3\sum_{s=1,2}\rho_{s,L}\rho_{s,R}&=&\frac{g_3}{4\pi}[(\partial_x \phi_1)^2-(\partial_x \theta_1)^2+(\partial_x \phi_2)^2-(\partial_x \theta_2)^2] \nonumber \\
&=&\frac{g_3}{4\pi}[(\partial_x \phi_+)^2-(\partial_x \theta_+)^2+(\partial_x \phi_-)^2-(\partial_x \theta_-)^2]
\eea

\bea
g_4\sum_{s\neq s'\in\{1,2\}}\rho_{s,L}\rho_{s',R}&=&\frac{g_4}{2\pi}[(\partial_x\phi_1)(\partial_2 \phi_2)-(\partial_x \theta_1)(\partial_x \theta_2)] \nonumber \\
&=&\frac{g_4}{4\pi}[(\partial_x \phi_+)^2-(\partial_x \theta_+)^2-(\partial_x \phi_-)^2+(\partial_x \theta_-)^2]
\eea

\bea
g_5\sum_{s=1,2}(\rho_{s,L}^2+\rho_{s,R}^2)&=&\frac{g_5}{2\pi}(\partial_x \phi_1)^2+(\partial_x \theta_1)^2+(\partial_x \phi_2)^2+(\partial_x \theta_2)^2] \nonumber \\
&=&\frac{g_5}{2\pi}[(\partial_x \phi_+)^2+(\partial_x \theta_+)^2+(\partial_x \phi_-)^2+(\partial_x \theta_-)^2]
\eea
It is then straightforward to arrive at
\bea
K_+&=&\sqrt{\frac{v_f+\frac{1}{2\pi}(2g_5+g_3+g_4)}{v_f+\frac{1}{2\pi}(2g_5-g_3-g_4)}} \nonumber \\
K_-&=&\sqrt{\frac{v_f+\frac{1}{2\pi}(2g_5+g_3-g_4)}{v_f+\frac{1}{2\pi}(2g_5-g_3+g_4)}}
\eea
To make pair-hopping interaction $g$ relevant, we require that $K_-<1$, which is equivalent to $g_3<g_4$.

\section{Appendix D. Estimation of $g$}

\begin{figure}[t]
	\centering
	\includegraphics[width=0.4\textwidth]{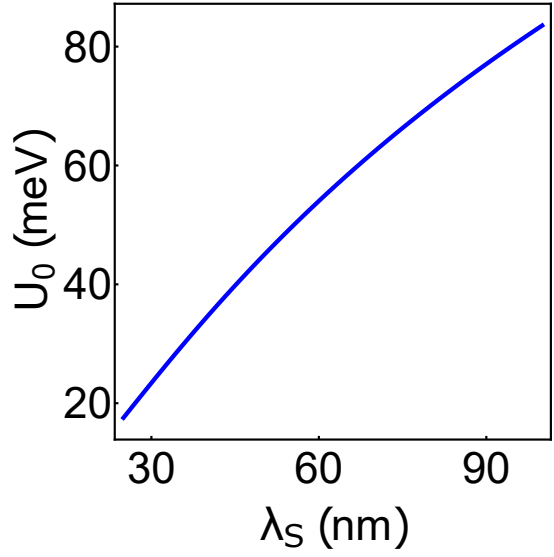}
	\caption{The dependence of interaction strength $U_0$ on the screening length $\lambda_s$.}
	\label{Fig: screening}
\end{figure}

In this section, we give a rough estimation of the magnitude of $g$. Our starting point is the standard Coulomb interaction,
\bea
U_C=\sum_{i,j\in\{1,2\}}\int d^3 {\bf r} d^3 {\bf r}'\rho_i({\bf r})V({\bf r-r}')\rho_j({\bf r}')
\eea
where $i$ and $j$ are the channel index. In the cylindrical geometry, ${\bf r}=(r\cos\theta,r\sin\theta,x)$. The screened Coulomb potential is defined as
\bea
V({\bf r})=\frac{e^2}{4\pi\epsilon_0\epsilon}\frac{1}{|{\bf r}|}e^{-|{\bf r}|/\lambda_{s}}
\eea
where $\lambda_{s}$ is the screening length to be determined. A general electron operator in the cylindrical geometry can be written as follows:
\bea
c^{\dagger}_i({\bf r})=\frac{1}{L}f(r,\theta)\sum_k e^{ikx} c^{\dagger}_i(k)
\eea
where $f(r,\theta)$ is the radial and angular part of the wavefunction. $k$ is the momentum defined in the $x$ direction (along the nanowire). The density operator is then defined as
\bea
\rho_i({\bf r})=\frac{1}{L}|f(r,\theta)|^2\rho_i(x)=\frac{1}{L}|f(r,\theta)|^2\sum_q\rho_i(q)e^{iqx}.
\eea
Since we are only interested in the Fermi-arc states which are exponentially localized along the radial direction, we assume $r\ll x$ in the long nanowire limit, and find that
\bea
|{\bf r-r}'|=\sqrt{(r\cos\theta-r'\cos\theta')^2+(r\sin\theta-r'\sin\theta')^2+(x-x')^2}\approx |x-x'|.
\eea
Back to $U_C$, we find that
\bea
U_C&\approx&\frac{1}{L^2}\sum_{i,j}(\int rdrd\theta |f(r,\theta)|^2) (\int r'dr'd\theta' |f(r',\theta')|^2)\int dx\int dx'\sum_{q,q'}\rho_i(q)\rho_j(q')e^{iqx+iq'x'}V(x-x') \nonumber \\
&=& \frac{1}{L^2}\sum_{i,j}\int dx\int dx'\sum_{q,q'}\rho_i(q)\rho_j(q')e^{iqx+iq'x'}V(x-x') \nonumber \\
&=&\frac{1}{L^2}\sum_{i,j}\int dx\int d\tilde{x}\sum_{q,q'}\rho_i(q)\rho_j(q')e^{i(q+q')x}V(\tilde{x})e^{-iq'\tilde{x}} \nonumber \\
&=&\sum_{i,j}\int d\tilde{x}V(\tilde{x})\sum_{q}\rho_i(q)\rho_j(-q) \nonumber \\
&=&\sum_{i,j}\int d\tilde{x}V(\tilde{x})\int dx \rho_i(x)\rho_j(x).
\eea
Here we have applied the long wave-length limit: $\lim_{q'\rightarrow 0}e^{-iq'\tilde{x}}=1$. Therefore, the four fermion contact interaction in the Luttinger liquid theory corresponds to the zero-momentum component of Coulomb interaction. In other words, the following expression will give us a rough estimation of the interaction strength in our system,
\bea
U_0= 2\int_{x>0} dx \frac{e^2}{4\pi\epsilon_0\epsilon}\frac{1}{x}e^{-x/\lambda_{s}} \approx \int_{l_B}^{\lambda_{s}}dx \frac{e^2}{2\pi\epsilon_0\epsilon}\frac{1}{x}=\frac{e^2}{2\pi\epsilon_0\epsilon}\ln (\frac{\lambda_{s}}{l_B}),
\eea
where $l_B$ is the magnetic length. As we have demonstrated in Appendix B, for $r=50$ nm, the magnetic field for inducing $2\pi$ flux is around $1T$. This gives rise to a magnetic length of around $25$ nm. For Cd$_3$As$_2$, the static dielectric constant measured is $\epsilon\approx 36$ \cite{jay1977}. An accurate estimate for the screening length $\lambda_s$ is difficult in our nanowire system. 
Thus, we take $\lambda_s$ ranging from $10$ to $100$ nm and numerically plot the dependence of $U_0$ on the screening lengthin Fig. \ref{Fig: screening}. We find that $U_0$ will increase with $\lambda_s$ and is typically around dozens of meV.

\section{Appendix E. Refermionization and Klein factors}
Introducing Klein factors is necessary to guarantee the anti-commutation relations of fermionic operators before and after the bosonization process, especially in a multiple-channel Luttinger liquid system. Meanwhile, Klein factors act as ladder operators in the Fock space of fermions. In this section, we present a detailed discussion of the role of Klein factors in our theory \cite{cheng2011}. In particular, we focus on the Luther-Emery point when the model is exactly solvable. Recall that the standard Abelian bosonization takes the form:
\bea
\psi_{s,R}=\frac{\xi_{s,R}}{\sqrt{2\pi a_0}}e^{i\sqrt{\pi}(\phi_s-\theta_s)},\ \ \
\psi_{s,L}=\frac{\xi_{s,L}}{\sqrt{2\pi a_0}}e^{-i\sqrt{\pi}(\phi_s+\theta_s)},
\eea
where $a_0$ is the short-distance cut-off. Klein factors $\xi_{l,R/L}$ satisfy the following relations,
\bea
[N_{s,h},\xi^{\dagger}_{s',h'}]&=&\xi^{\dagger}_{s,h}\delta_{h,h'}\delta_{s,s'},
\ \ \
[N_{s,h},\xi_{s',h'}]=-\xi_{s,h}\delta_{h,h'}\delta_{s,s'}, \nonumber \\
\{\xi_{s,h},\xi_{s',h'}\}&=&0,\ \ \
\{\xi_{s,h},\xi^{\dagger}_{s',h'}\}=2\delta_{s,s'}\delta_{h,h'},\ \ \
\xi_{s,h}\xi^{\dagger}_{s,h}=1.\ \ \
\eea
Here $N_{s,h}$ is the number operators of fermions with channel index $s=1,2$ and chirality $h=R/L$. At the Luther-Emery point ($K_-=\frac{1}{2}$), we could map the pair-hopping interaction $g$ into a spinless p-wave pairing term via refermionization process:
\bea
\begin{tikzcd}[column sep=9pc]
	\psi^{\dagger}_{2,R}\psi^{\dagger}_{2,L}\psi_{1,R}\psi_{1,L} \arrow{r}{K_-=\frac{1}{2}} \arrow{d}{\text{Bosonization}} &
	\tilde{\psi}_R\tilde{\psi}_L \\
	\xi^{\dagger}_{2,R}\xi^{\dagger}_{2,L}\xi_{1,R}\xi_{1,L}e^{i2\sqrt{2\pi}\theta_-} \arrow{r}{\text{Field rescaling}} &
	F_RF_Le^{i2\sqrt{\pi}\tilde{\theta}} \arrow{u}{\text{Refermionization}}
\end{tikzcd}
\label{Eq: Refermionization}
\eea
As a result, we start from $H_-$ with $K_-=\frac{1}{2}$ and arrive at the following Hamiltonian that turns out to be the Kitaev model:
\bea
H_-=iv\int dx (\tilde{\psi}_R^{\dagger}\partial_x\tilde{\psi}_R-\tilde{\psi}_L^{\dagger}\partial_x\tilde{\psi}_L)+im\int dx (\tilde{\psi}_R\tilde{\psi}_L+\tilde{\psi}_R^{\dagger}\tilde{\psi}_L^{\dagger})
\eea
The new defined fermionic operators are
\bea
\tilde{\psi}_R=\frac{F_R}{\sqrt{2\pi a_0}}e^{i\sqrt{\pi}(\frac{\phi_-}{\sqrt{2}}-\sqrt{2}\theta_-)},\ \ \
\tilde{\psi}_L=\frac{F_L}{\sqrt{2\pi a_0}}e^{-i\sqrt{\pi}(\frac{\phi_-}{\sqrt{2}}+\sqrt{2}\theta_-)}
\eea
Following Ref. \cite{cheng2011,kraus2013}, a natural choice for defining the new Klein factors would be
\bea
F_R=f_R\xi^{\dagger}_{2,R}\xi_{1,R},\ \ \ F_L=f_L\xi^{\dagger}_{2,L}\xi_{1,L}.
\eea
In the above definition, we have introduced an additional ``Majorana" operator $f_{R/L}$ to guarantee the anti-commutation relation between $F_R$ and $F_L$, which commutes with $N_{s,h}$ and thus acts trivially on the Fock space. One can easily check that
\bea
[N_{s,h},F_{h'}]=(-1)^sF_h\delta_{h,h'},\ \ \ \text{where }s,s'\in\{1,2\},\ h,h'\in\{R,L\}.
\eea

\section{Appendix F. Majorana zero mode at Luther-Emery point}
We perform the Bogoliubov-de Gennes (BdG) transformation by defining the BdG basis
\bea
\tilde{\Psi}=\begin{pmatrix}
	\tilde{\psi}_R \\
	\tilde{\psi}_L^\dagger
\end{pmatrix}
\eea
Then $H_-$ becomes a 1D massive Dirac Hamiltonian under the BdG basis:
\bea
H_-=\int dx \tilde{\Psi}^{\dagger}(iv\partial_x\sigma_z-m\sigma_y)\tilde{\Psi}
\eea
This Hamiltonian can be easily solved for boundary zero mode by imposing open boundary condition $\tilde{\Psi}(x<0)=0$, and we obtain a zero-mode solution that is exponentially localized at $x=0$ \cite{cheng2011}:
\bea
\alpha_1(x)=\sqrt{\frac{m}{v}}e^{-\frac{mx}{v}}\begin{pmatrix}
	1 \\
	1
\end{pmatrix}
\tilde{\Psi}(x)=\sqrt{\frac{m}{v}}e^{-\frac{mx}{v}}[\tilde{\psi}_R(x)+\tilde{\psi}_L^\dagger(x)]
\eea
In the bosonic language, we show in the main text that the open boundary condition requires
\bea
\phi_-=(2n_--n_+)\sqrt{\frac{\pi}{2}}.
\eea
where $n_{\pm}\in \mathbb{Z}$. $n_+$ can be viewed as a c-number in our discussion, and we are free to take $n_+=0$ for simplicity. This bosonic boundary condition simply implies
\bea
\tilde{\psi}_R(x=0)=\tilde{\psi}_L (x=0)
\eea
This immediately leads to
\bea
\alpha_1=\alpha_1^{\dagger}
\eea
at $x=0$, and verified $\alpha_1$ as a Majorana zero mode. At Luther-Emery point, it is clear that the bosonized form of $\alpha_1$ (Eq. [6] in the main text) should inherit the Klein factor $F_h$ from the bosonization formula of $\tilde{\psi}_h$.

\section{Appendix G. Single impurity problem}

In this section, we briefly discuss the stability of Majorana physics in the presence of $C_4$-breaking disorders. For simplicity, we will focus on the case with a single impurity at $x=0$. With such impurity, a left-moving electron in channel 1 can be either forward-scattered into a left-moving electron in channel 2 (denoted as $T_1$), or be backscattered into a right-moving electron in channel 2 (denoted as $T_2$). Since the scattering physics happens only at the impurity site $x=0$, we are free to integrate out all bosonic fields except for the ones at $x=0$. The system then reduces to a 0+1 dimensional field theory with,
\bea
T_1&=&t_1\psi^{\dagger}_{1,L}(x=0)\psi_{2,L}(x=0)+h.c.\sim t_1\cos\sqrt{2\pi}(\phi_-+\theta_-), \nonumber\\
T_2&=&t_2\psi^{\dagger}_{1,L}(x=0)\psi_{2,R}(x=0)+h.c.\sim t_2\cos\sqrt{2\pi}(\phi_++\theta_-).
\eea
The scaling dimensions of $T_1$ and $T_2$ are
\bea
\Delta(T_1)=\frac{1}{2}(K_-+\frac{1}{K_-}),\ \ \ \Delta(T_2)=\frac{1}{2}(K_-+\frac{1}{K_+}).
\eea
Under renormalization group (RG) process, $T_1$ is always marginal or irrelevant. $T_2$ is irrelevant when $\Delta(T_2)>1$, or equivalently
\bea
K_+<\frac{1}{2-K_-}\leq 1
\label{Eq: Disorder irrelevant}
\eea
The second inequality is true when the pair-hopping interaction is relevant under RG ($K_-<1$). Therefore, when Eq. \ref{Eq: Disorder irrelevant} holds, $C_4$-breaking impurity is irrelevant and the $C_4$-protected Majorana physics is stable.

\section{Appendix H. Emergent $Z_M$ fermion parity symmetry from a $Z_N$ symmetry}
In this section, we hope to establish the equivalence between a discrete $Z_N$ symmetry with the fermion parity symmetry. As an example, we first start from the model of DSM nanowire with a $C_4$ rotation ($Z_4$) symmetry. According to the angular momentum representations, $C_4$ acts on the dual bosonic fields as,
\bea
C_4
\begin{pmatrix}
	\phi_+ \\
	\phi_- \\
	\theta_+ \\
	\theta_- \\
\end{pmatrix}
=\begin{pmatrix}
	\phi_+ \\
	\phi_- \\
	\theta_+ - \frac{1}{2}\sqrt{\frac{\pi}{2}} \\
	\theta_- +\sqrt{\frac{\pi}{2}} \\
\end{pmatrix}
\eea
Thus, $C_4$ symmetry also interchanges $|\theta_-=\pm\frac{1}{2}\sqrt{\frac{\pi}{2}}\rangle$, which indicates that the degenerate ground states $|\pm\rangle$ are also eigen-states of $C_4$ symmetry operator. This proves the equivalence between $C_4$ symmetry and $Z_2$ parity in the low-energy sector.

Now let us consider a two-channel 1D system with a $Z_N$ symmetry. The fermionic operators of different channels belong to distinct representations of the $Z_N$ symmetry,
\bea
Z_N\ \psi_{1,L/R}\ Z_N^{-1}&=&e^{i\frac{2\pi}{N}J}\psi_{1,L/R} \nonumber \\
Z_N\ \psi_{2,L/R}\ Z_N^{-1}&=&e^{i\frac{2\pi}{N}(J+k)}\psi_{2,L/R}
\eea
where $0\leq J<N$, $0<k<N$ and $J,k\in \mathbb{Z}$. A general multi-fermion interaction takes the form,
\bea
U=\psi_{1,L}^{m_1}\psi_{1,R}^{n_1}\psi_{2,L}^{m_2}\psi_{2,R}^{n_2}.
\eea
where $m_{1,2},n_{1,2}\in \mathbb{Z}$. If $m_{i}$ and $n_i$ take negative value, this represents $(\psi_{i,L}^{\dagger})^{|m_i|}$ and $(\psi_{i,R}^{\dagger})^{|n_i|}$ in $U$. Define $L_1=m_1+n_1$ and $L_2=m_2+n_2$, then $L_1=-L_2=M$ is required by charge conservation symmetry. Under $Z_N$ symmetry, we find that
\bea
Z_N U Z_N^{-1}=e^{i\frac{2\pi}{N}[J(L_1+L_2)+kL_2]}U=e^{-i\frac{2\pi}{N}kM}U
\eea
Therefore, to make $U$ invariant under $Z_N$, we require
\bea
kM\equiv 0\ \text{mod }N\ \ \ \Rightarrow\ \ \ M=\frac{LCM(N,k)}{k}.
\label{Eq: Fermion parity & Z_N}
\eea
where $LCM(N,k)$ is the least common multiple of $N$ and $k$. Physically, $M$ is the number of electrons that tunnel from channel 1 to channel 2. When $U$ is large and relevant under RG, the electron number in each channel is well-defined modulo $M$, which signals an emergent $Z_M$ fermion parity symmetry. In the following, we will discuss two special situations of Eq. \ref{Eq: Fermion parity & Z_N}:
\begin{itemize}
	\item {\bf When $LCM(N,k)=N$, $M=\frac{N}{k}\in \mathbb{Z}$. Then $Z_N$ symmetry leads to an emergent $Z_{\frac{N}{k}}$ fermion parity.} For the DSM nanowire model in the main text, we have $N=4$ and  $k=\frac{3}{2}-(-\frac{1}{2})=2$. Thus, $M=LCM(4,2)/2=2$, which corresponds to an emergent $Z_2$ fermion parity.
	\item {\bf When $LCM(N,k)=Nk$, $M=N$. Then $Z_N$ symmetry leads to an emergent $Z_N$ fermion parity.} As an example, we consider a system where $\psi_{1,L/R}$ carries angular momentum $\frac{1}{2}$, while $\psi_{2,L/R}$ carries angular momentum $-\frac{1}{2}$. For $N=4$, now we have $k=\frac{1}{2}-(-\frac{1}{2})=1$ and $M=LCM(4,1)/1=4$. This leads to an emergent $Z_4$ fermion parity.
\end{itemize}

\section{Appendix I. Emergent $Z_2$ fermion parity from 2D point groups}
Given a $Z_N$ group and its representations, we have shown in the previous section that an $Z_M$ fermion parity symmetry will emerge for each representation even when the total number of electrons is conserved, which is summarized in Eq. \ref{Eq: Fermion parity & Z_N}. In this section, we hope to look at the problem from a different perspective: Given a group $G$, we are wondering what representations of $G$ could lead to an emergent $Z_2$ fermion parity symmetry ($M=2$). In particular, we hope to go beyond $Z_N$ symmetry and explore a general symmetry group $G$. We first propose three criteria that $G$ should satisfy:

\begin{enumerate}
	\item[{\bf (a)}] {\bf $G$ is ``on-site".} In a quasi-1D nanowire system growing along $x$ direction, only symmetry operations that act trivially on $x$ can be viewed as on-site symmetries. This implies that our target symmetry groups should be 2D point groups.
	\item[{\bf (b)}] {\bf $Z_2$ is a subgroup of $G$.} When this condition is satisfied, there exist two inequivalent 1D irreps with basis functions $|J\rangle$ and $|J+k\rangle$ such that $M=LCM(N,k)/k=2$.
	\item[{\bf (c)}] {\bf There exists an interaction $g$ such that: (i) $g$ describes pair-hopping process between $Z_2=\pm 1$ representations; (ii) $g$ respects every symmetry operation in $G$.}
	\label{Condition: general requirements for G}
\end{enumerate}

There are in total 10 point groups in 2D lattice systems: $C_n$ and $D_n$, where $n=1,2,3,4,6$. Here $C_n$ group contains a n-fold rotation symmetry as the group generator. $D_n$ group contains an additional in-plane mirror operation compared to $C_n$ group. Notice that $C_1=\{e\}$ is a trivial identity group, while $D_1=\{e,M\}$ where $M$ is a mirror operation.

\subsection{I1. $C_n$ group}
A $C_n$ group is just a $Z_n$ group. $C_n$ groups have only 1D irreps labeled by angular momentum $J$, where $J$ is integer (half-integer) for spinless (spinful) electrons. Starting from Eq. \ref{Eq: Fermion parity & Z_N}, we have $M=2$ for $Z_2$ fermion parity. Thus, $LCM(n,k)=2k$ and we have $n=2k$, which leads to
\bea
e^{i\frac{2\pi}{n}(J'-J)}=e^{i\frac{2\pi}{n}k}=-1
\label{Eq: Z_2 criterion of C_n}
\eea
In the following, we give a detailed discussion of all $C_n$ groups and check their compatibility with Eq. \ref{Eq: Z_2 criterion of C_n}. These results are summarized in Table \ref{Table:C_n group}.
\begin{itemize}
	\item {\bf $C_2$ group (spinless)} - There are two inequivalent 1D irreps: $|J=0\rangle$ and $|J=1\rangle$, which satisfies Eq. \ref{Eq: Z_2 criterion of C_n}
	\item {\bf $C_2$ group (spinful)} - The 1D irreps are $|J=\frac{1}{2}\rangle$ and $|J=-\frac{1}{2}\rangle$ and satisfies Eq. \ref{Eq: Z_2 criterion of C_n}.
	\item {\bf $C_3$ group (spinless)} - The 1D irreps are $|J=-1\rangle$, $|J=0\rangle$ and $J=1\rangle$. None of them satisfy Eq. \ref{Eq: Z_2 criterion of C_n}.
	\item {\bf $C_3$ group (spinful)} - The 1D irreps are $|J=-\frac{1}{2}\rangle$, $|J=\frac{1}{2}\rangle$ and $J=\frac{3}{2}\rangle$. None of them satisfy Eq. \ref{Eq: Z_2 criterion of C_n}.
	\item {\bf $C_4$ group (spinless)} - The 1D irreps are $|J=0\rangle$, $|J=\pm 1\rangle$, and $|J=2\rangle$. It is easy to check that both $\{|J=0\rangle,|J=2\rangle\}$ and $\{|J=1\rangle,|J=-1\rangle\}$ satisfy Eq. \ref{Eq: Z_2 criterion of C_n}.
	\item {\bf $C_4$ group (spinful)} - The 1D irreps are $|J=\pm \frac{1}{2}\rangle$ and $|J=\pm \frac{3}{2}\rangle$. It is easy to check that both $\{|J=\frac{1}{2}\rangle,|J=-\frac{3}{2}\rangle\}$ and $\{|J=-\frac{1}{2}\rangle,|J=\frac{3}{2}\rangle\}$ satisfy Eq. \ref{Eq: Z_2 criterion of C_n}.
	\item {\bf $C_6$ group (spinless)} - The 1D irreps are $|J=0\rangle$, $|J=\pm 1\rangle$, $|J=\pm2\rangle$, and $|J=3\rangle$. It is easy to check that $\{|J=0\rangle,|J=3\rangle\}$, $\{|J=-1\rangle,|J=2\rangle\}$ and $\{|J=-2\rangle,|J=1\rangle\}$ satisfy Eq. \ref{Eq: Z_2 criterion of C_n}.
	\item {\bf $C_6$ group (spinful)} - The 1D irreps are $|J=\pm\frac{1}{2}\rangle$, $|J=\pm \frac{3}{2}\rangle$, and $|J=\pm \frac{5}{2}\rangle$. It is easy to check that $\{|J=\frac{1}{2}\rangle,|J=-\frac{5}{2}\rangle\}$, $\{|J=\frac{3}{2}\rangle,|J=-\frac{3}{2}\rangle\}$ and $\{|J=\frac{5}{2}\rangle,|J=-\frac{1}{2}\rangle\}$ satisfy Eq. \ref{Eq: Z_2 criterion of C_n}.
\end{itemize}

\begin{table}
	\begin{tabular}{cccccc}
		\toprule[1.5pt] \\[-1em]
		& ${\bf C_2}\ $ & ${\bf C_3}\ $ & ${\bf C_4}\ $ & ${\bf C_6}\ $ \\
		\midrule \hline \\
		Spinless: &  $|J=0,1 \rangle\ $ & N/A\ \  & $|J=0,2\rangle\ \ $  & $|J=0,3\rangle$ \\[0.5em]
		&  & & $|J=\pm 1\rangle$ & $|J=1,-2\rangle$ \\[0.5em]
		&  & & & $|J=2,-1\rangle$ \\
		\\[-0.5em] \hline \\[-0.5em]
		Spinful:  & $|J=\pm \frac{1}{2} \rangle\ $ & N/A\ \  & $|J=\frac{1}{2},-\frac{3}{2}\rangle\ \ $  & $|J=\frac{1}{2},-\frac{5}{2}\rangle$ \\[0.5em]
		&  & & $|J=-\frac{1}{2},\frac{3}{2}\rangle$ & $|J=\pm\frac{3}{2}\rangle$ \\[0.5em]
		&  & & & $|J=\frac{5}{2},-\frac{1}{2}\rangle$ \\
		\\
		\bottomrule[1.5pt]
	\end{tabular}
	\caption{$C_n$ group}
	\label{Table:C_n group}
\end{table}

\subsection{I2. $D_n$ group}

With an additional mirror operation, irreps of $D_n$ group can be characterized by either angular momentum $J$ or mirror eigenvalue $m$, while we will continue to use $J$ to label representations. For $n>2$, $D_n$ group is non-Abelian due to the non-commutation between rotation symmetry $C_n$ and mirror symmetry $M$, giving rise to 2D irreps. In the following, we first derive a general condition (Eq. \ref{Eq: Condition_Dn}) for $D_n$ group and pair-hopping interactions, and then apply this condition to each $D_n$ group. Given a state $|J\rangle$, we have
\bea
M|J\rangle=c|-J\rangle
\eea
Here $c=1$ for spinless electrons and $c=i$ for spinful electrons. In the basis $\Psi_{L/R}=(\psi_{J,L/R},\psi_{-J,L/R})^T$, the symmetry operations are
\bea
C_n=e^{i\frac{2\pi}{n}J\sigma_z},\ \ \ M=\sigma_x
\eea
The pair-hopping interaction between different irreps is
\bea
V_{pair}=\psi_{J,L}^{\dagger}\psi_{J,R}^{\dagger}\psi_{-J,L}\psi_{-J,R}+h.c= \Psi^{\dagger}_{L}\sigma_+\Psi_{L}\Psi^{\dagger}_{R}\sigma_+\Psi_{R}+h.c.
\eea
where $\sigma_{\pm}=\sigma_x\pm i\sigma_y$ are the pseudo-spin Pauli matrices. It is important to check the compatibility between $V_{pair}$ and symmetry operations. Notice that
\bea
M\sigma_+M^{-1}&=&\sigma_- \nonumber \\
C_n\sigma_+ C_n^{-1}&=&e^{i\frac{4\pi}{n}J}\sigma_+
\eea
Therefore,
\bea
MV_{pair}M^{-1}&=&V_{pair} \nonumber \\
C_nV_{pair}C_n^{-1}&=&e^{i\frac{8\pi}{n}J}V_{pair}
\eea
Under $n$-fold rotation symmetry, $V_{pair}$ is invariant when the following conditions are satisfied,
\bea
4J&=&kn,\ \ \ k\in\mathbb{Z} \nonumber \\
2J&<&n
\label{Eq: Condition_Dn}
\eea
The second condition in Eq. \ref{Eq: Condition_Dn} guarantees that $|J\rangle$ and $|-J\rangle$ are different states.\\

Meanwhile, we can reformulate our argument from mirror symmetry eigenstates
\bea
|m_{\pm}\rangle=\frac{1}{\sqrt{2}}(|J\rangle\pm|-J\rangle)
\eea
where $M|m_{\pm}=\pm|m_{\pm}\rangle$. In this case, we could define the pair-hopping interaction under a new basis $\tilde{\Psi}_{L/R}=(\psi_{m_+,L/R},\psi_{m_-,L/R})^T$,
\bea
\tilde{V}_{pair}=\tilde{\Psi}^{\dagger}_{L}\sigma_+\tilde{\Psi}_{L}\tilde{\Psi}^{\dagger}_{R}\sigma_+\tilde{\Psi}_{R}+h.c.
\eea
In this case, the symmetry operations are
\bea
\tilde{C}_n=e^{i\frac{2\pi}{n}J\sigma_x}, \ \ \ \tilde{M}=c\sigma_z.
\eea
It is easy to see that $V_{pair}$ must satisfy Eq. \ref{Eq: Condition_Dn} to respect symmetries. The results of our detailed analysis below is summarized in Table \ref{Table:D_n group}.

\begin{itemize}
	\item {\bf $D_1$ group (spinless)} - $D_1$ group has no rotational symmetry operation, so we use mirror eigenvalue to label 1D irreps of $D_1$ group as $|m=1\rangle$ and $|m=-1\rangle$, which satisfies our requirement for hosting Majorana.
	\item {\bf $D_1$ group (spinful)} - The $D_1$ double group is similar to its spinless version. The 1D irreps that will work are $|m=i\rangle$ and $|m=-i\rangle$.
	\item {\bf $D_2$ group (spinless)} - This is similar to the case of spinless $C_2$ group, and the 1D irreps are $|J=0\rangle$ and $|J=1\rangle$.
	\item {\bf $D_2$ group (spinful)} - Different from the spinless $D_2$ group, $|J=\frac{1}{2}\rangle$ and $|J=-\frac{1}{2}\rangle$ form a 2D irrep since $M|J=\frac{1}{2}\rangle=i|J=-\frac{1}{2}\rangle$. It is easy to see that $|J=\frac{1}{2}\rangle$ and $|J=-\frac{1}{2}\rangle$ satisfy Eq. \ref{Eq: Condition_Dn}.
	\item {\bf $D_3$ group (spinless $\&$ spinful)} - For $n=3$, Eq. \ref{Eq: Condition_Dn} becomes: $4J=3k$ and $J<\frac{3}{2}$. It is easy to check that for both spinless ($J$ is integer) and spinful ($J$ is half-integer) cases, there is no $J$ satisfying both equations.
	\item {\bf $D_4$ group (spinless $\&$ spinful)} - For $n=4$, it is easy to check that Eq. \ref{Eq: Condition_Dn} becomes: $J=k$ and $J<2$. Thus, $J$ can only take integer values $J=1$, which corresponds to the spinless case.
	\item {\bf $D_6$ group (spinless $\&$ spinful)} - For $n=6$, Eq. \ref{Eq: Condition_Dn} is: $2J=3k$ and $J<3$. The only solution is $J=\frac{3}{2}$.
\end{itemize}

\begin{table}
	\begin{tabular}{cccccc}
		\toprule[1.5pt] \\[-1em]
		& ${\bf D_1}$ & ${\bf D_{2}}\ $ & ${\bf D_{3}}\ $ & ${\bf D_{4}}\ $ & ${\bf D_{6}}\ $ \\
		\midrule \hline \\
		Spinless: & $|M=\pm 1\rangle\ $ & $|J=0,1 \rangle\ $ & N/A\ \  & $|J=\pm 1\rangle\ \ $  & N/A \\
		\\[-0.5em] \hline \\[-0.5em]
		Spinful: & $|M=\pm i\rangle\ $ & $|J=\pm \frac{1}{2} \rangle\ $ & N/A\ \  & N/A\ \  & $|J=\pm \frac{3}{2}\rangle$ \\
		\\
		\bottomrule[1.5pt]
	\end{tabular}
	\caption{$D_{n}$ group}
	\label{Table:D_n group}
\end{table}

\section{Appendix J. Diagonalizing $H_-$ with mode expansion}

In this appendix, we introduce mode expansion technique to explicitly diagonalize the anti-bonding Hamiltonian $H_-$. The mode expansion offers us a powerful weapon to attack the electron tunneling problem. We introduce,
\bea
\phi_-(x)&=&\sqrt{\frac{\pi}{2}}(n_{\phi}+\frac{1}{2})+\sqrt{\frac{\pi}{2}}\frac{\Delta n x}{L}+\frac{1}{\sqrt{\pi}}\sum_{k=1}^{\infty}\sqrt{\frac{1}{K_-k}}\sin\frac{\pi k x}{L}(a_k+a_k^{\dagger}) \nonumber \\
\theta_-(x)&=&\sqrt{\frac{\pi}{2}}(n_{\theta}+\frac{1}{2})+\sqrt{\frac{\pi}{2}}\theta_0-i\frac{1}{\sqrt{\pi}}\sum_{k=1}^{\infty}\sqrt{\frac{K_-}{k}}\cos\frac{\pi k x}{L}(a_k-a_k^{\dagger})
\label{Eq: Mode expansion}
\eea
with $[\Delta n,\theta_0]=i$ and $[a_k,a_{k'}^{\dagger}]=\delta_{k,k'}$. Here, we have expanded the cosine term in $H_-$ around its minimum as $4\pi g[\theta_- - (n_{\theta}+\frac{1}{2})\sqrt{\frac{\pi}{2}}]^2$. Plug in the mode expansion (Eq. \ref{Eq: Mode expansion}) to the Hamiltonian and after some tedious calculations, we obtain that
\bea
\frac{v}{2}K_-\int dx (\partial_x\phi_-)^2&=&\frac{v\pi}{4L}(\Delta n)^2+\sum_k\frac{v\pi k}{4L}[a_ka_k^{\dagger}+a_k^{\dagger}a_k+(a_k)^2+(a_k^{\dagger})^2] \nonumber \\
\frac{v}{2}\frac{1}{K_-}\int dx (\partial_x\theta_-)^2&=&-\sum_k\frac{v\pi k}{4L}[-a_ka_k^{\dagger}-a_k^{\dagger}a_k+(a_k)^2+(a_k^{\dagger})^2] \nonumber \\
4\pi g[\theta_- - (n_{\theta}+\frac{1}{2})\sqrt{\frac{\pi}{2}}]^2&=&2gL\pi^2\theta_0^2-\sum_k\frac{2gLK_-}{k}[-a_ka_k^{\dagger}-a_k^{\dagger}a_k+(a_k)^2+(a_k^{\dagger})^2]
\eea
Therefore,
\bea
H_-=\frac{v\pi}{4L}(\Delta n)^2+2gL\pi^2\theta_0^2+\sum_k \frac{A_k}{2}[a_ka_k^{\dagger}+a_k^{\dagger}a_k]-\sum_k \frac{B_k}{2}[(a_k)^2+(a_k^{\dagger})^2].
\eea
With
\bea
A_k&=&\frac{v\pi k}{L}+\frac{4gLK_-}{k} \nonumber \\
B_k&=&\frac{4gLK_-}{k}
\eea
Introduce the following Bogoliubov transformations
\bea
\Delta n&=&(\frac{8g\pi L^2}{v})^{\frac{1}{4}}\frac{d+d^{\dagger}}{\sqrt{2}} \nonumber \\
\theta_0&=&(\frac{8g\pi L^2}{v})^{-\frac{1}{4}}\frac{d-d^{\dagger}}{i\sqrt{2}} \nonumber \\
a_k&=&C_k \alpha_k+D_k \alpha^{\dagger}_k
\eea
where $C_k=\sqrt{\frac{1}{2}(\frac{A_k}{E_k}+1)}$ and $D_k=\sqrt{\frac{1}{2}(\frac{A_k}{E_k}-1)}$. The Bogoliubov quasi-particles satisfy commutation relations as $[d,d^{\dagger}]=1$ and $[\alpha_k,\alpha_{k'}^{\dagger}]=\delta_{k,k'}$. The energy function $E_k$ is now given by $E_k=\sqrt{A_k^2-B_k^2}$, and it is closely related to the energy dispersion of the system
\bea
H_-=\sqrt{\frac{vg\pi^3}{2}}d^{\dagger}d+\sum_k E_k \alpha^{\dagger}_k\alpha_k
\eea
Let us have a closer look at the form of $E_k$. By defining $q=\frac{\pi k }{L}$, we find a very nice form
\bea
E_k=\sqrt{v^2q^2+\Delta^2}
\eea
where $\Delta=\sqrt{8\pi vgK_-}$ physically denotes the energy gap formed in the anti-bonding sector, which originates from pair-hopping interaction $g$.

\section{Appendix K. Transition amplitude of electron tunneling problem}
When an electron of $s=1$ tunnels into the system, it changes the sign of both total fermion parity $P_{tot}$ and $P^{(2)}$. Therefore, the original ground state $|0\rangle=|N,p\rangle$ is changed to $|1\rangle=|N+1,-p\rangle$ via this tunneling process, with $p\in{\pm1}$. On the other hand, the total charge $Q_{+}$ and the charge in channel $s=1$ are given by
\bea
Q_{+}&=&\sqrt{\frac{2}{\pi}}[\phi_+(L)-\phi_+(0)], \nonumber \\
Q_1&=&\sqrt{\frac{1}{\pi}}[\phi_1(L)-\phi_1(0)]=\frac{1}{2}(Q_+ + Q_-)
\eea
Since $\Delta Q_+=\Delta Q_1=1$ during this single electron tunneling, we find that $\Delta Q_+=\Delta Q_-=1$, or equivalently
\bea
\delta\phi_+=\delta\phi_-=\sqrt{\frac{\pi}{2}}.
\eea
where we have defined $\delta\phi_{\pm}=\phi_{\pm}(L)-\phi_{\pm}(0)$. Here it is convenient to parameterize $\Delta\phi_{\pm}(x)=\sqrt{\frac{\pi}{2}}(1-\frac{x}{L})$. The two ground states can be connected via a unitary transformation $|1\rangle=e^{-i\eta}|0\rangle$  \cite{keselman2015}, where $\eta=\eta_+ + \eta_-$ and
\bea
\eta_+&=&\int dx \Delta\phi_+\partial_x \theta_+ \nonumber \\
\eta_-&=&\int dx \Delta\phi_-\partial_x \theta_- .
\eea
With the fact that $[e^{-i\eta_{\pm}},\phi_{\pm}(x)]=\Delta\phi_{\pm}(x)e^{-i\eta_{\pm}}$, it is easy to show that $e^{-i\eta_{\pm}}\phi_{\pm}(x)e^{i\eta_{\pm}}=\phi_{\pm}(x)+\Delta \phi_{\pm}(x)$.

Next, we would like to calculate the tunneling amplitude
\bea
\langle 1 | \psi_{1,R}^{\dagger} |0\rangle&=&\langle 0|e^{i\eta}e^{-i\sqrt{\pi}(\phi_1-\theta_1)}|0\rangle \nonumber \\
&=&\langle e^{i\eta_+}e^{-i\sqrt{\frac{\pi}{2}}(\phi_+-\theta_+)}\rangle_+ \times \langle e^{i\eta_-}e^{-i\sqrt{\frac{\pi}{2}}(\phi_--\theta_-)}\rangle_- \nonumber \\
&=&T_+T_-
\eea
Let us first focus on the anti-bonding part,
\bea
T_-&=&\langle e^{i\eta_-}e^{-i\sqrt{\frac{\pi}{2}}(\phi_--\theta_-)}\rangle_- \nonumber \\
&=&e^{-\frac{1}{2}\langle [\eta_- -\sqrt{\frac{\pi}{2}}(\phi_--\theta_-)]^2\rangle_-}
\label{Eq: T_-}
\eea

\subsection{K1. Contribution from $T_-$}

With the mode expansion in Eq. \ref{Eq: Mode expansion}, let us move on to calculate $T_-$. For $\delta_-$
\bea
\eta_-&=&\int dx \Delta\phi_-\partial_x \theta_- \nonumber \\
&=&\frac{i}{\sqrt{2}}\sum_k\sqrt{\frac{K}{k}}\frac{\pi k}{L}(a_k-a_k^{\dagger})\int_0^L dx (1-\frac{x}{L})\sin\frac{\pi k x}{L} \nonumber \\
&=&i\sqrt{\frac{K}{2}}\sum_k\frac{a_k-a_k^{\dagger}}{\sqrt{k}}
\eea
Notice that in $T_-=e^{-F_-}$, the imaginary part in $F_-=\frac{1}{2}\langle [\eta_- -\sqrt{\frac{\pi}{2}}(\phi_--\theta_-)]^2\rangle_-$ only contributes to a phase factor, while its amplitude originates from its real part. By ignoring the imaginary part, we find that
\bea
F_-=\frac{1}{2}\langle [\eta_-+\sqrt{\frac{\pi}{2}}\theta_-]^2\rangle_-+\frac{\pi}{4}\langle\phi_-^2\rangle_-
\eea
To start with
\bea
\eta_-+\sqrt{\frac{\pi}{2}}\theta_-=i\sum_k\sqrt\frac{K_-}{2k}(a_k-a_k^{\dagger})(1-\cos\frac{\pi k x}{L})+\frac{\pi}{2}(n_{\theta}+\theta_0+\frac{1}{2})
\eea
In the following discussion, we will ignore the $\theta_0$ and $\Delta n$ part, which will give rise to a universal contribution to $T_-$ as
\bea
e^{-\frac{\pi^2}{8}[(n_{\theta}+\frac{1}{2})^2+(n_{\phi}+\frac{1}{2})^2]}\times e^{-\frac{\pi^2}{16L}(\sqrt{\frac{v}{8g\pi}}+\sqrt{\frac{8g\pi}{v}}x^2)}
\eea
where the second term becomes unity as $L\rightarrow \infty$. Then
\bea
\langle [\eta_-+\sqrt{\frac{\pi}{2}}\theta_-]^2\rangle_-&=&-\sum_{k,k'}\frac{K_-}{2}\frac{1}{\sqrt{k k'}}(1-\cos\frac{\pi k x}{L})(1-\cos\frac{\pi k' x}{L})(a_k-a_k^{\dagger})(a_{k'}-a_{k'}^{\dagger}) \nonumber \\
&=&\sum_k (C_k-D_k)^2\frac{K_-}{2k}(1-\cos\frac{\pi k x}{L})^2
\eea
where we have used $\langle \alpha^{\dagger}_k\alpha_k\rangle=0$. Let us define $\kappa=\xi^{-1}=\frac{\Delta}{v}$, and the correlation length is defined as
\bea
\xi=\frac{v}{\Delta}=\sqrt{\frac{v}{8\pi g K_-}}.
\label{Eq: Correlation length}
\eea
Then we have
\bea
A_k&=&vq+\frac{\Delta^2}{2qv}=v(q+\frac{\kappa^2}{2q}) \nonumber \\
B_k&=&v\frac{\kappa^2}{2q}
\eea
In the strong coupling limit $\Delta\rightarrow \infty$, $\kappa\gg q$. Then $A_k\approx B_k\approx v\frac{\kappa^2}{2q}$ and $E_k\approx v\kappa$. We find that
\bea
(C_k-D_k)^2\approx \frac{q}{\kappa}
\eea
By taking $L\rightarrow \infty$,
\bea
\frac{1}{2}\langle [\eta_-+\sqrt{\frac{\pi}{2}}\theta_-]^2\rangle_-&=&\frac{K_-}{4\kappa}\int dq (1-\cos qx)^2 \times e^{-\epsilon q} \nonumber \\
&=&\frac{3K_-}{8}\frac{\xi}{\epsilon}
\eea
where we have introduced a decay factor $e^{-\epsilon q}$ as the short distance divergence, and $\epsilon$ is the short distance cut-off. Therefore, this part of $F_-$ does not have spatial dependence. For the second part of $F_-$,
\bea
\frac{\pi}{4}\langle\phi_-^2\rangle_-&=&\frac{1}{4 K_-}\sum_{k,k'}\sqrt{\frac{1}{k k'}}\sin \frac{\pi k x}{L}\sin \frac{\pi k' x}{L}(a_k+a_k^{\dagger})(a_{k'}+a_{k'}^{\dagger}) \times e^{-\epsilon q} \nonumber \\
&=&\frac{1}{4 K_-}\sum_{k}\frac{(C_k+D_k)^2}{k}\sin^2 \frac{\pi k x}{L} \times e^{-\epsilon q} \nonumber \\
&=&\frac{\kappa}{4 K_-}\int dq \frac{\sin^2 qx}{q^2} \times e^{-\epsilon q} \nonumber \\
&=&\frac{\kappa}{4 K_-}[x\arctan\frac{2x}{\epsilon}-\frac{1}{4}\epsilon\log(\frac{4x^2+\epsilon^2}{\epsilon^2})] \nonumber \\
&=&\frac{\pi}{8K_-}\frac{x}{\xi}-\frac{\epsilon}{8K_-\xi}\log\frac{2x}{\epsilon}
\eea
Finally, we arrive at the following important result,
\bea
T_-\sim(\frac{2x}{\epsilon})^{\frac{\epsilon}{8K_-\xi}}\times e^{-\frac{\pi}{8K_-}\frac{x}{\xi}}
\eea

\subsection{K2. Contribution from $T_+$}
Since the bonding part is gapless, we expect it only gives a power-law correction to $T$. To see this, we notice that the essential difference between bonding and anti-bonding physics lies in the existence of $g$. By setting $g=0$ in anti-bonding sector, one should be able to recover the physics in the bonding sector. In this limit, we find that
\bea
\Delta=0,\ E_k=A_k=vq,\ B_k=0.
\eea
This gives rise to $C_k=1$ and $D_k=0$. Then
\bea
\frac{1}{2}\langle [\eta_++\sqrt{\frac{\pi}{2}}\theta_+]^2\rangle_+&=&\frac{K_+}{4}\int dq \frac{(1-\cos qx)^2}{q} \times e^{-\epsilon q} \nonumber \\
&=&\frac{K_+}{4}[\log\frac{x^2+\epsilon^2}{\epsilon^2}-\frac{1}{4}\log\frac{4x^2+\epsilon^2}{\epsilon^2}]
\eea
and
\bea
\frac{\pi}{4}\langle\phi_+^2\rangle_+&=&\frac{1}{4 K_+}\int dq \frac{\sin^2 qx}{q} \times e^{-\epsilon q} \nonumber \\
&=&\frac{1}{16K_+}\log\frac{4x^2+\epsilon^2}{\epsilon^2}
\eea
Then we arrive at
\bea
T_+=(\frac{\epsilon^2}{x^2+\epsilon^2})^{\frac{K_+}{4}}(\frac{4x^2+\epsilon^2}{\epsilon^2})^{\frac{1}{16}(K_+-\frac{1}{K_+})}
\eea
Together with the anti-bonding contribution, we conclude that the single electron tunneling amplitude is given by
\bea
\langle 1 | \psi_{1,R}^{\dagger} |0\rangle = {\cal N}(x,\epsilon,K_{\pm}) e^{-\frac{\pi}{8K_-}\frac{x}{\xi}}.
\label{Eq: T exponential}
\eea
where ${\cal N}(x,\epsilon,K_{\pm})$ denotes the power-law corrections.

\section{Appendix L. General Formalism of DEBC method}
In this appendix, we briefly review the general formalism of ``delayed evaluation of boundary condition" (DEBC) method \cite{oshikawa2006}, and how this method can help us to evaluate scaling dimension of operators with given boundary conditions. To start with, let us consider N quantum wires with different Luttinger parameter $K_i$ ($i=1,2,...,N$) meet at $x=0$ and form a junction. The bosonized Hamiltonian is given by
\bea
H=\sum_{i=1}^N\int dx [K_i(\partial_x\phi_i)^2+\frac{1}{K_i}(\partial_x\theta_i)^2 ]
\eea
where abelian bosonization is defined as
\bea
\psi_{i,R}&\sim& e^{i\sqrt{\pi}\chi_{i,R}}=e^{i\sqrt{\pi}(\phi_i-\theta_i)} \nonumber \\
\psi_{i,L}&\sim& e^{-i\sqrt{\pi}\chi_{i,L}}=e^{-i\sqrt{\pi}(\phi_i+\theta_i)}
\eea
It is convenient to rescale the dual fields to make the system effectively non-interacting,
\bea
\tilde{\phi}_i&=&\sqrt{K_i}\phi_i,\ \tilde{\theta}_i=\frac{1}{\sqrt{K_i}}\theta_i \nonumber \\
\tilde{\chi}_{i,R}&=&\tilde{\phi}_i-\tilde{\theta}_i\ \tilde{\chi}_{i,L}=\tilde{\phi}_i+\tilde{\theta}_i
\label{Eq: Fields rescaling}
\eea
The boundary condition is demonstrated by
\bea
\vec{\chi}_R={\cal O} \vec{\chi}_L
\label{Eq: Def of O}
\eea
where $\vec{\chi}_{R/L}=(\tilde{\chi}_{1,R/L},\tilde{\chi}_{2,R/L},...,\tilde{\chi}_{N,R/L})^T$. ${\cal O}$ is an orthogonal matrix with ${\cal O}{\cal O}^T=1$, which contains key information of boundary condition. In other words, {\bf a particular choice of ${\cal O}$ determines certain type of conformal invariant boundary condition, which corresponds to certain RG fixed point in the tunneling phase diagram.}

To be specific, perturbations of a boundary fixed point come from tunneling or backscattering process, which generally can be written as
\bea
g_{m,n}\sim e^{i\sqrt{\pi}({\bf m}\cdot \vec{\phi}+{\bf n}\cdot \vec{\theta})}
\eea
where $\vec{\phi}=(\tilde{\phi}_1,\tilde{\phi}_2,...,\tilde{\phi}_N)^T$ and $\vec{\theta}=(\tilde{\theta}_1,\tilde{\theta}_2,...,\tilde{\theta}_N)^T$. With the definition of ${\cal O}$ matrix, the boundary condition can be easily implemented in the $\tilde{\chi}_{R,L}$ basis. Thus,
\bea
g_{m,n}&\sim & e^{i\frac{\sqrt{\pi}}{2}[({\bf m}+{\bf n})\cdot \vec{\chi}_R+({\bf m}-{\bf n})\cdot \vec{\chi}_L]} \nonumber \\
&=& e^{i\frac{\sqrt{\pi}}{2}[{\cal O}^T({\bf m}+{\bf n})+({\bf m}-{\bf n})]^T \vec{\chi}_L} \nonumber \\
\label{Eq: g_(m,n)}
\eea
Its scaling dimension is
\bea
\Delta[g_{m,n}]=\frac{1}{8}|{\cal O}^T({\bf m+n})+({\bf m-n})|^2
\label{Eq: Scaling dimension}
\eea
The specific form of ${\cal O}$ depends on the boundary condition we impose.

\subsection{L1. Orthogonality of ${\cal O}$ matrix}
The orthogonality condition turns out to be an important and strong constraint to ${\cal O}$, which can be proved as follows. Consider an operator $g_{A,B}=e^{i\frac{\sqrt{\pi}}{2}({\bf A}\cdot\vec{\chi}_R+{\bf B}\cdot\vec{\chi}_L)}$, with $\bf A$ and $\bf B$ both $N$-component vectors. From Eq. \ref{Eq: Def of O}, $\vec{\chi}_L={\cal O}^{-1}\vec{\chi}_R$. Then
\bea
g_{A,B}&=&e^{i\frac{\sqrt{\pi}}{2}({\bf A}^T{\cal O}+{\bf B}^T)\vec{\chi}_L} \nonumber \\
&=&e^{i\frac{\sqrt{\pi}}{2}({\bf A}^T+{\bf B}^T{\cal O}^{-1})\vec{\chi}_R}
\eea
In terms of $\vec{\chi}_L$, the scaling dimension of $g_{A,B}$ is
\bea
\Delta(g_{A,B})&=&\frac{1}{8}|{\bf A}^T{\cal O}+{\bf B}^T|^2 \nonumber \\
&=&\frac{1}{8}[{\cal O}^T({\bf A}{\bf A}^T){\cal O}+{\bf B}{\bf B}^T+{\cal O}^T{\bf A}{\bf B}^T+{\bf B}{\bf A}^T{\cal O}] \nonumber \\
&=&\frac{1}{8}[|{\bf A}|^2{\cal O}^T{\cal O}+|{\bf B}|^2+({\cal O}^{T}+{\cal O}){\bf A}\cdot {\bf B}]
\label{Eq: l.h.s}
\eea
In terms of $\vec{\chi}_R$, we find that
\bea
\Delta(g_{A,B})&=&\frac{1}{8}|{\bf A}^T+{\bf B}^T{\cal O}^{-1}|^2 \nonumber \\
&=&\frac{1}{8}[|{\bf A}|^2+({\cal O O^T})^{-1}|{\bf B}|^2+({\cal O}+{\cal O^T})^{-1}){\bf A}\cdot {\bf B}].
\label{Eq: r.h.s}
\eea
To make Eq. \ref{Eq: l.h.s} and Eq. \ref{Eq: r.h.s} self-consistent, orthogonality condition ${\cal O}{\cal O}^T=1$ must be satisfied.

\subsection{L2. Application of DEBC to 2LL/LL junction}
Now, we are ready to apply the DEBC method to our 2LL/LL junction setup. The central task here is to identify ${\cal O}$ matrix that characterizes physical RG fixed point. Across the junction, current conservation requires
\bea
J_{x<0}+J_{x>0}=0
\eea
Here $J_{x<0}=-\sqrt{\frac{1}{\pi}}\partial_t\phi_0$ and $J_{x>0}=-\sqrt{\frac{1}{\pi}}\partial_t(\phi_1+\phi_2)=\sqrt{\frac{2}{\pi}}\partial_t\phi_+$. The rescaling rule of fields can be defined similar  to Eq. \ref{Eq: Fields rescaling}, with $i=0,\pm$. Then the current conservation condition becomes
\bea
\partial_t(\frac{1}{\sqrt{K_0}}\tilde{\phi}_0+\sqrt{\frac{2}{K_+}}\tilde{\phi}_+)=0
\label{Eq: Current conservation}
\eea
Other information of ${\cal O}$ depends on the details of boundary conditions. Once we obtain ${\cal O}$, it is important to write down possible tunneling or backscattering operators that could perturb the fixed point.\\

(i) \textit{Single electron tunneling},
\bea
T_{ij}^s=\psi_{i,R}^{\dagger}\psi_{j,L}+h.c.\sim\cos\sqrt{\pi}[(\phi_i+\phi_j)-(\theta_i-\theta_j)],
\eea
where $i\neq j\in\{0,1,2\}$. However, electron tunneling process should conserve angular momentum. Let us consider a case where electron from LL lead can hop to both $i=1$ and $i=2$. But inter-channel single electron tunneling events ($T^s_{12}$ and $T^s_{21}$) in 2LL are strictly forbidden due to the angular momentum conservation.
\\

(ii) \textit{Single electron backscattering},
\bea
B_i=\psi_{i,L}^{\dagger}\psi_{i,R}+h.c.\sim \cos2\sqrt{\pi}\phi_i,
\label{Eq:Backscattering cosine}
\eea
with $i\in\{0,1,2\}$.
\\

(iii) \textit{Pair electron tunneling},
\bea
T_{ij}^p=\psi_{i,R}^{\dagger}\psi_{i,L}^{\dagger}\psi_{j,R}\psi_{j,L}+h.c.\sim \cos2\sqrt{\pi}(\theta_i-\theta_j)
\eea
with $i<j\in\{0,1,2\}$. Notice that the pair electron tunneling is essentially different from $T^2_{ij}$.

\section{Appendix M. Phase diagram of Majorana-free 2LL/LL junction}

In this section, let us first ignore pair hopping interaction between $i=1$ and $i=2$ that pins $\theta_-$. Then the system is very similar to a conventional ``Y" junction composed of three independent quantum wires, while rMZM does NOT show up at the junction. However, an important difference from previous work \cite{oshikawa2006,hou2012} comes from the constraint of angular momentum conservation. In particular, electron tunneling between $i=1$ and $i=2$ is forbidden to conserve angular momentum. For our purpose, we will focus on the situation where $K_-<1$.

To identify a fixed point, an effective way is to assume that some electron operators have vanishing scaling dimensions. This ansatz immediately leads to a corresponding ${\cal O}$ matrix. With this ${\cal O}$ matrix, we first check its orthogonality condition to verify its validity. A stable fixed point is confirmed when all possible perturbations are found to be irrelevant, with the help of Eq. \ref{Eq: Scaling dimension}. Below, we first use ``disconnecting fixed point" as a detailed example to demonstrate how DEBC method extracts the information of a fixed point. Then we will exhaust all physical RG fixed points and discuss their stability problem.

\subsection{M1. Disconnecting fixed point: an tutorial of DEBC method}
The disconnecting fixed point is characterized by the vanishing scaling dimension of all the backscattering processes $B_{0,1,2}$. Thanks to the current conservation, the vanishing scaling dimension of two backscattering operators will automatically leads to zero scaling dimension of the remaining backscattering operator. Physically, this simply means that if channel $i=0$ and $i=1$ are both disconnected, $i=2$ is also forced to be disconnected at the junction as no electron can flow from $i=0,1$ to $i=2$.

We first define $\vec{\chi}_{R/L}=(\tilde{\chi}_{0,R/L},\tilde{\chi}_{+,R/L},\tilde{\chi}_{-,R/L})^T$, following the convention in Eq. \ref{Eq: Fields rescaling}. The backscattering operators are now:
\bea
B_0 &\sim & \cos \sqrt{\pi}(\frac{\tilde{\chi}_{0,R}+\tilde{\chi}_{0,L}}{\sqrt{K_0}}) \nonumber \\
B_1&\sim & \cos \sqrt{\pi}[\frac{\tilde{\chi}_{+,R}+\tilde{\chi}_{+,L}}{\sqrt{2K_+}}+\frac{\tilde{\chi}_{-,R}+\tilde{\chi}_{-,L}}{\sqrt{2K_-}}] \nonumber \\
B_2&\sim & \cos \sqrt{\pi}[\frac{\tilde{\chi}_{+,R}+\tilde{\chi}_{+,L}}{\sqrt{2K_+}}-\frac{\tilde{\chi}_{-,R}+\tilde{\chi}_{-,L}}{\sqrt{2K_-}}]
\eea

Let us re-examine Eq. \ref{Eq: g_(m,n)} and Eq. \ref{Eq: Scaling dimension}, the only
way to make $\Delta[g_{m,n}]=0$ is to impose
\bea
({\bf m+n})\cdot \vec{\chi}_R+({\bf m-n})\cdot \vec{\chi}_L=0.
\label{Eq: Vanishing scaling dimension}
\eea

Therefore, we impose Eq. \ref{Eq: Vanishing scaling dimension} to $B_0$ and $B_1$. Together with the current conservation condition, we arrive at
\bea
\frac{\tilde{\chi}_{0,R}+\tilde{\chi}_{0,L}}{\sqrt{K_0}}&=&0 \nonumber \\
\frac{\tilde{\chi}_{+,R}+\tilde{\chi}_{+,L}}{\sqrt{2K_+}}+\frac{\tilde{\chi}_{-,R}+\tilde{\chi}_{-,L}}{\sqrt{2K_-}}&=&0 \nonumber \\
\frac{\tilde{\chi}_{0,R}+\tilde{\chi}_{0,L}}{2\sqrt{K_0}}+\frac{\tilde{\chi}_{+,R}+\tilde{\chi}_{+,L}}{\sqrt{2K_+}}&=&0
\eea
It is quite easy to see the solution of the above equations as
\bea
\tilde{\chi}_{w,R}=-\tilde{\chi}_{w,L},\ \ \ \forall w\in\{0,1,2\},
\label{Eq: Solution of DFP}
\eea
which immediately leads to the rotation matrix ${\cal O}_D$ that characterizes disconnecting fixed point according to its definition (Eq. \ref{Eq: Def of O}):
\bea
{\cal O}_D=\begin{pmatrix}
	-1 & 0 & 0 \\
	0 & -1 & 0 \\
	0 & 0 & -1 \\
\end{pmatrix}.
\label{DFP O matrix}
\eea
It is easy to see that ${\cal O}_D$ is an orthogonal matrix. Physically, Eq. \ref{Eq: Solution of DFP} implies
\bea
\phi_0=\phi_+=\phi_-=0.
\label{Eq: DFP boundary condition}
\eea
which is recognized as the disconnecting boundary condition at the junction. The scaling dimensions of electron operators can be calculated via Eq. \ref{Eq: Scaling dimension},
\bea
\Delta(T^s_{10})&=&\Delta(T^s_{01})=\Delta(T^s_{20})=\Delta(T^s_{02})=\frac{1}{4}(2K_0+K_++K_-) \nonumber \\
\Delta(T^p_{01})&=&2K_0+K_++K_- \nonumber \\
\Delta(T^p_{12})&=&4K_- \nonumber \\
\Delta(T^p_{20})&=&2K_0+K_++K_- \nonumber \\
\Delta(B_{0})&=&0 \nonumber \\
\Delta(B_{1})&=&0 \nonumber \\
\Delta(B_{2})&=&0.
\label{Eq: Delta of DFP}
\eea
When all tunneling terms are irrelevant, we require
\bea
K_0&>&\frac{(4-K_-)-K_+}{2} \nonumber \\
K_-&>&\frac{1}{4}
\label{Eq: DFP stability}
\eea
This condition explicitly characterizes the region in the tunneling phase diagram where DFP is stable.

\subsection{M2. Pair-tunneling fixed point}
At the pair-tunneling fixed point, two-electron pair-tunneling process $T_{ij}^P$ is dominating, and
\bea
{\cal O}_{P}=\begin{pmatrix}
	-1+\frac{4K_0}{2K_0+K_+} & -\frac{2\sqrt{2K_0K_+}}{2K_0+K_+} & 0 \\
	-\frac{2\sqrt{2K_0K_+}}{2K_0+K_+} & 1-\frac{4K_0}{2K_0+K_+} & 0 \\
	0 & 0 & 1 \\
\end{pmatrix}.
\label{Eq: PTFP O matrix}
\eea
The ${\cal O}_P$ matrix is obtained similarly to ${\cal O}_D$ matrix in Eq. \ref{DFP O matrix} in the last section. The orthogonality of ${\cal O}_D$ can be checked easily. From ${\cal O}_P$, it is easy to tell that
\bea
\phi_0=-\sqrt{2}\phi_+,\ \theta_+=\sqrt{2}\theta_0, \ \theta_-=0.
\eea
The scaling dimensions at this fixed point are
\bea
\Delta(T^s_{10})&=&\Delta(T^s_{01})=\Delta(T^s_{20})=\Delta(T^s_{02})=\frac{1}{4}(\frac{1}{2K_0+K_+}+\frac{1}{K_-}) \nonumber \\
\Delta(T^p_{01})&=&0 \nonumber \\
\Delta(T^p_{12})&=&0 \nonumber \\
\Delta(T^p_{20})&=&0 \nonumber \\
\Delta(B_{0})&=&\frac{4}{2K_0+K_+} \nonumber \\
\Delta(B_{1})&=&\frac{1}{2K_0+K_+}+\frac{1}{K_-} \nonumber \\
\Delta(B_{2})&=&\frac{1}{2K_0+K_+}+\frac{1}{K_-}.
\eea
Pair-tunneling fixed point is stable when
\bea
\frac{1}{2K_0+K_+}+\frac{1}{K_-}&>&4 \nonumber \\
2K_0+K_+&<&4
\label{Eq: PTFP stability}
\eea
Thus
\bea
K_0<\frac{C_p-K_+}{2}
\eea
where $C_p=4$ if $K_-\leq\frac{1}{4}$ and $C_p=\text{min}\{4,(4-\frac{1}{K_-})^{-1}\}$ if $K_->\frac{1}{4}$.

\subsection{M3. $\chi_+$ fixed point}
$\chi_+$ fixed point is defined so that $T_{10}^s$ and $T^s_{02}$ are the only terms whose scaling dimension is vanishing, and thus dominate the junction physics. We find that
\bea
{\cal O}_{+}=\frac{1}{1+K_-(2K_0+K_+)}
\begin{pmatrix}
	-1+K_-(2K_0-K_+) & -2K_-\sqrt{2K_0K_+} & 2\sqrt{2K_0K_-} \\
	-2K_-\sqrt{2K_0K_+} & -1-K_-(2K_0-K_+) & -2\sqrt{K_+K_-} \\
	-2\sqrt{2K_0K_-} & 2\sqrt{K_+K_-} & -1+K_-(2K_0+K_+) \\
\end{pmatrix}.
\label{Eq: chi_+ O matrix}
\eea
Despite its complicated form, one can still check that ${\cal O}_+{\cal O}_+^T=1$ is satisfied. From ${\cal O}_+$, this simply means
\bea
\phi_0=-\sqrt{2}\phi_+=\sqrt{2}\theta_-,\ \phi_-=\theta_+-\sqrt{2}\theta_0.
\eea
The scaling dimensions at this fixed point are
\bea
\Delta(T^s_{10})&=&\Delta(T^s_{02})=0 \nonumber \\
\Delta(T^s_{01})&=&\Delta(T^s_{20})=\frac{2K_0+K_++K_-}{1+K_-(2K_0+K_+)} \nonumber \\
\Delta(T^p_{01})&=&\frac{2K_0+K_++K_-}{1+K_-(2K_0+K_+)} \nonumber \\
\Delta(T^p_{12})&=&\frac{4K_-}{1+K_-(2K_0+K_+)} \nonumber \\
\Delta(T^p_{20})&=&\frac{2K_0+K_++K_-}{1+K_-(2K_0+K_+)} \nonumber \\
\Delta(B_{0})&=&\frac{4K_-}{1+K_-(2K_0+K_+)} \nonumber \\
\Delta(B_{1})&=&\frac{2K_0+K_++K_-}{1+K_-(2K_0+K_+)} \nonumber \\
\Delta(B_{2})&=&\frac{2K_0+K_++K_-}{1+K_-(2K_0+K_+)}.
\eea
$\chi_+$ fixed point is stable when
\bea
[1-(2K_0+K_+)](1-K_-)&<&0 \nonumber \\
K_-[4-(2K_0+K_+)]&>&1
\label{Eq: chi_+ stability}
\eea
Since $K_-<1$, we find that
\bea
\frac{1-K_+}{2}&<&K_0<\frac{(4-\frac{1}{K_-})-K_+}{2} \nonumber \\
\text{and  } K_-&>&\frac{1}{3}
\eea

\subsection{M4. $\chi_-$ fixed point}
When $T_{01}^s$ and $T^s_{20}$ are dominating, we arrive at the $\chi_-$ fixed point, with
\bea
{\cal O}_{-}=\frac{1}{1+K_-(2K_0+K_+)}
\begin{pmatrix}
	-1+K_-(2K_0-K_+) & -2K_-\sqrt{2K_0K_+} & -2\sqrt{2K_0K_-} \\
	-2K_-\sqrt{2K_0K_+} & -1-K_-(2K_0-K_+) & 2\sqrt{K_+K_-} \\
	2\sqrt{2K_0K_-} & -2\sqrt{K_+K_-} & -1+K_-(2K_0+K_+) \\
\end{pmatrix}.
\label{Eq: chi_- O matrix}
\eea
and
\bea
\phi_0=-\sqrt{2}\phi_+=-\sqrt{2}\theta_-,\ \phi_-=-\theta_++\sqrt{2}\theta_0.
\eea
Notice the difference and similarity between ${\cal O}_{+}$ and ${\cal O}_-$. It is easy to check that $\chi_-$ fixed point shares the same scaling dimensions of $T^p_{ij}$ and $B_i$ with those of $\chi_+$ fixed point. The only change in scaling dimension is that
\bea
\Delta(T^s_{10})&=&\Delta(T^s_{02})=\frac{2K_0+K_++K_-}{1+K_-(2K_0+K_+)} \nonumber \\
\Delta(T^s_{01})&=&\Delta(T^s_{20})=0
\eea
Therefore, when Eq. \ref{Eq: chi_+ stability} is satisfied, both $\chi_+$ and $\chi_-$ fixed points are stable. In other words, $\chi_+$ and $\chi_-$ fixed points coexist, and we expect an interesting intermediate unstable fixed point which characterizes the phase transition between these two fixed points.

\subsection{M5. $A_0$ fixed point}
Asymmetric fixed point $A_i$ takes place when channel $i$ is disconnected to the rest of the junction, while the other two channels are fully connected via electron tunneling process. For $A_0$ fixed point, $B_0$ and $T_{12}^p$ will have vanishing scaling dimension, with
\bea
{\cal O}_{A_0}=
\begin{pmatrix}
	-1 & 0 & 0 \\
	0 & -1 & 0 \\
	0 & 0 & 1 \\
\end{pmatrix}.
\label{Eq: A_0 O matrix}
\eea
and
\bea
\phi_0=\phi_+=\theta_-=0.
\eea
${\cal O}_{A_0}$ is obviously orthogonal. The scaling dimensions at this fixed point are
\bea
\Delta(T^s_{10})&=&\Delta(T^s_{01})=\Delta(T^s_{02})=\Delta(T^s_{20})=\frac{1+K_-(2K_0+K_+)}{4K_-} \nonumber \\
\Delta(T^p_{01})&=&2K_0+K_+ \nonumber \\
\Delta(T^p_{12})&=&0 \nonumber \\
\Delta(T^p_{20})&=&2K_0+K_+ \nonumber \\
\Delta(B_{0})&=&0 \nonumber \\
\Delta(B_{1})&=&\frac{1}{K_-} \nonumber \\
\Delta(B_{2})&=&\frac{1}{K_-}.
\eea
For $K_-<1$, $B_{1,2}$ are always irrelevant under RG. Thus, $A_0$ fixed point is stable when
\bea
2K_0+K_+&>&4-\frac{1}{K_-} \nonumber \\
2K_0+K_+&>&1
\eea
Thus,
\bea
K_0>\frac{C_{A_0}-K_+}{2}
\eea
where $C_{A_0}=\text{max}\{1,4-\frac{1}{K_-}\}$.

\subsection{M6. $A_1$ fixed point}
For $A_1$ fixed point, we require $T^{s}_{02}$, $T^s_{20}$, $T^p_{20}$ and $B_1$ possessing zero scaling dimension. It turns out that these conditions are compatible with each other, leading to rotation matrix
\bea
{\cal O}_{A_1}=\frac{1}{2K_0+K_++K_-}
\begin{pmatrix}
	2K_0-K_+-K_- & -2\sqrt{2K_0K_+} & 2\sqrt{2K_0K_-} \\
	-2\sqrt{2K_0K_+} & -2K_0+K_+-K_- & -2\sqrt{K_+K_-} \\
	2\sqrt{2K_0K_-} & -2\sqrt{K_+K_-} & -2K_0-K_++K_- \\
\end{pmatrix}.
\label{Eq: A_1 O matrix}
\eea
Despite its complicated form, one can still check that ${\cal O}_{A_1}{\cal O}_{A_1}^T=1$ is satisfied. We find that
\bea
\phi_0=\sqrt{2}\phi_-,\ \phi_+=-\phi_-, \theta_-+\sqrt{2}\theta_0=\theta_+,
\eea
and
\bea
\Delta(T^s_{10})&=&\Delta(T^s_{01})=\frac{1+K_-(2K_0+K_+)}{2K_0+K_++K_-} \nonumber \\
\Delta(T^s_{02})&=&\Delta(T^s_{20})=0 \nonumber \\
\Delta(T^p_{01})&=&4(\frac{1}{K_-}+\frac{1}{2K_0+K_+})^{-1} \nonumber \\
\Delta(T^p_{12})&=&4(\frac{1}{K_-}+\frac{1}{2K_0+K_+})^{-1} \nonumber \\
\Delta(T^p_{20})&=&0 \nonumber \\
\Delta(B_{0})&=&\frac{4}{2K_0+K_++K_-} \nonumber \\
\Delta(B_{1})&=&0 \nonumber \\
\Delta(B_{2})&=&\frac{4}{2K_0+K_++K_-}.
\eea
Thus, $A_1$ fixed point is stable when
\bea
[1-(2K_0+K_+)](1-K_-)&>&0 \nonumber \\
\frac{1}{K_-}+\frac{1}{2K_0+K_+}&<&4 \nonumber \\
2K_0+K_++K_-&<&4
\eea
Thus,
\bea
\frac{(4-\frac{1}{K_-})^{-1}-K_+}{2}&<&K_0<\frac{1-K_+}{2} \nonumber \\
K_-&>&\frac{1}{4}
\eea

\subsection{M7. $A_2$ fixed point}
For $A_2$ fixed point, the terms $T^{s}_{01}$, $T^s_{10}$, $T^p_{01}$ and $B_2$ can carry zero scaling dimension simultaneously, giving rise to
\bea
{\cal O}_{A_1}=\frac{1}{2K_0+K_++K_-}
\begin{pmatrix}
	2K_0-K_+-K_- & -2\sqrt{2K_0K_+} & -2\sqrt{2K_0K_-} \\
	-2\sqrt{2K_0K_+} & -2K_0+K_+-K_- & 2\sqrt{K_+K_-} \\
	-2\sqrt{2K_0K_-} & 2\sqrt{K_+K_-} & -2K_0-K_++K_- \\
\end{pmatrix}.
\label{Eq: A_2 O matrix}
\eea
and
\bea
\phi_0=-\sqrt{2}\phi_-,\ \phi_+=\phi_-, \theta_-+\theta_+=\sqrt{2}\theta_0.
\eea
Then
\bea
\Delta(T^s_{10})&=&\Delta(T^s_{01})=0 \nonumber \\
\Delta(T^s_{02})&=&\Delta(T^s_{20})=\frac{1+K_-(2K_0+K_+)}{2K_0+K_++K_-} \nonumber \\
\Delta(T^p_{01})&=&0 \nonumber \\
\Delta(T^p_{12})&=&4(\frac{1}{K_-}+\frac{1}{2K_0+K_+})^{-1} \nonumber \\
\Delta(T^p_{20})&=&4(\frac{1}{K_-}+\frac{1}{2K_0+K_+})^{-1} \nonumber \\
\Delta(B_{0})&=&\frac{4}{2K_0+K_++K_-} \nonumber \\
\Delta(B_{1})&=&\frac{4}{2K_0+K_++K_-} \nonumber \\
\Delta(B_{2})&=&0.
\eea
Thus, $A_2$ fixed point is stable when
\bea
[1-(2K_0+K_+)](1-K_-)&>&0 \nonumber \\
\frac{1}{K_-}+\frac{1}{2K_0+K_+}&<&4 \nonumber \\
2K_0+K_++K_-&<&4
\eea
It is easy to see that $A_1$ and $A_2$ fixed points coexist in the same parameter region.

\subsection{M8. Majorana-free phase diagram without bulk pair-hopping interaction}
\begin{figure}[t]
	\centering
	\includegraphics[width=0.9\textwidth]{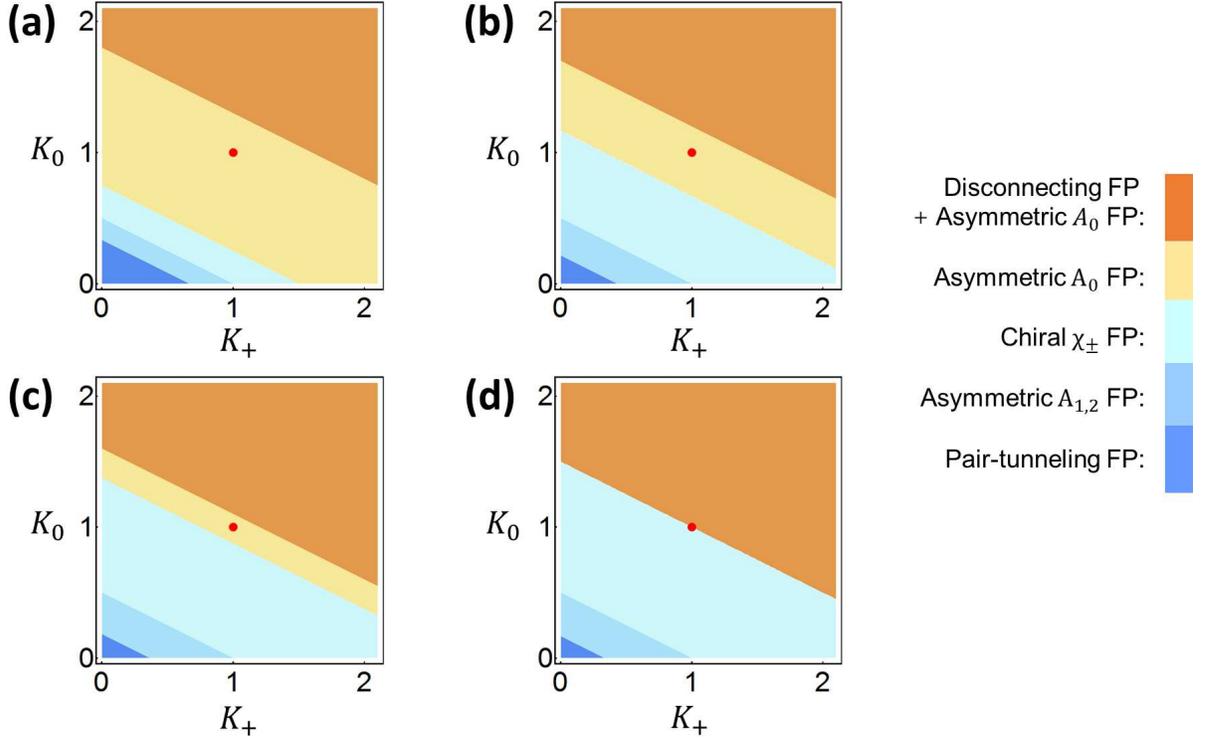}
	\caption{Majorana-free phase diagram with $K_-$ value: (a) $K_-=0.4$, (b) $K_-=0.6$, (c) $K_-=0.8$, (d) $K_-=1$ }
	\label{Fig: PD without pair hopping}
\end{figure}

Now we are ready to map out the phase diagram of DSM nanowire junction without pair-hopping interaction (Majorana-free). As shown in Fig. \ref{Fig: PD without pair hopping}, the phase diagrams are plotted for a fixed $K_-$ value in: (a) $K_-=0.4$, (b) $K_-=0.6$, (c) $K_-=0.8$, (d) $K_-=1$. Remarkably, the junction is at stable asymmetric $A_0$ fixed point in the ``free fermion" limit with ${K_0=K_+=1}$ (red dot), where wire $i=0$ is completely disconnected from wire $i=1,2$. In this weakly interacting region with $K_-<1$, the two-terminal conductance across the junction should be zero.

\section{Appendix N. Phase diagram of 2LL/LL junction with rMZM}

When bulk pair-hopping interaction is incorporated between channel $i=1$ and $i=2$, there exists a rMZM bound state at the 2LL/LL junction. Pair-hopping interaction $g$ pins $\theta_-$ to a classical value $\sqrt{\pi}n_{\theta_-}$, with $n_{\theta_-}\in \mathbb{Z}$. Interestingly, it is possible that certain boundary condition pins $\phi_-$ to $\sqrt{\pi}n_{\phi_-}x$ ($n_{\phi_-}\in\mathbb{Z}$) simultaneously at $x=0$, even though $\theta_-$ and $\phi_-$ do NOT commute with each other. Such competition between bulk interaction and boundary condition requires promoting both $n_{\phi_-}$ and $n_{\theta_-}$ to integer-valued operators to recover the non-commutativity between $\phi_-$ and $\theta_-$ at $x=0$. With this subtlety, we briefly summarize our strategy to study the critical behaviors here: \\

(1) For a generic tunneling/backscattering operator $t=\cos\sqrt{\pi}[(m_0\phi_0+m_+\phi_++m_-\phi_-)+(n_0\theta_0+n_+\theta_++n_-\theta_-)]$. The pinning of $\theta_-$ is equivalent to setting $n_-=0$ in $t$ in RG calculation. \\

(2) With $n_-=0$, we force $\Delta(t)$ (the scaling dimension of $t$) to be zero to arrive at a possible fixed point $F_t$, which is characterized by an ${\cal O}_t$ matrix. {\bf $F_t$ is a physical fixed point with $\Delta(t)=0$, only if orthogonality condition ${\cal O}_t^T{\cal O}_t=1$ is still lsatisfied. Otherwise, $F_t$ is not physical and does not show up in the phase diagram.} \\

(3) Write down the explicit boundary conditions of a physical fixed point, and check that whether the boundary condition forces the pinning of $\phi_-$. {\bf If pinning of $\phi_-$ is not required by the boundary condition, any operator $t$ with $m_-\neq 0$ is forced to vanish due to strong $\phi_-$ fluctuations and does not enter the phase diagram.}\\

(4) Use ${\cal O}_t$ to calculate the remaining perturbation terms, and obtain the stability condition of $F_t$.

\subsection{N1. Why $\chi_+$ fixed point is not physical in 2LL/LL junction with rMZM?}

To start with, we apply the above criterion to $\chi_+$ fixed point, and explicitly show that in the presence of pair hopping interaction $g$, the rotation matrix of $\chi_+$ fixed point ${\cal O}_+$ is modified to a non-orthogonal form ${\cal O}_+^{(g)}$. Recall that $\chi_+$ fixed point forces trivial scaling dimensions of $T_{10}^{s}$ and $T^s_{02}$. Let us first write down
\bea
T^s_{10}&\sim& \cos \sqrt{\pi}[\frac{\tilde{\chi}_{0,L}+\tilde{\chi}_{0,R}}{2\sqrt{K_0}}+\frac{\sqrt{K_0}}{2}(\tilde{\chi}_{0,L}-\tilde{\chi}_{0,R})+\frac{\tilde{\chi}_{+,L}+\tilde{\chi}_{+,R}}{2\sqrt{2K_+}}-\frac{\sqrt{K_+}}{2\sqrt{2}}(\tilde{\chi}_{+,L}-\tilde{\chi}_{+,R})+\frac{\tilde{\chi}_{-,L}+\tilde{\chi}_{-,R}}{2\sqrt{2K_-}}] \nonumber \\
T^s_{02}&\sim& \cos \sqrt{\pi} [\frac{\tilde{\chi}_{0,L}+\tilde{\chi}_{0,R}}{2\sqrt{K_0}}-\frac{\sqrt{K_0}}{2}(\tilde{\chi}_{0,L}-\tilde{\chi}_{0,R})+\frac{\tilde{\chi}_{+,L}+\tilde{\chi}_{+,R}}{2\sqrt{2K_+}}+\frac{\sqrt{K_+}}{2\sqrt{2}}(\tilde{\chi}_{+,L}-\tilde{\chi}_{+,R})-\frac{\tilde{\chi}_{-,L}+\tilde{\chi}_{-,R}}{2\sqrt{2K_-}}]
\eea
where we have ignored the constant part originated from the pinning of $\theta_-$. Together with the current conservation condition, the existence of $\chi_+$ fixed point requires the following equations to be satisfied
\bea
\frac{\tilde{\chi}_{0,R}+\tilde{\chi}_{0,L}}{\sqrt{2K_0}}+\frac{\tilde{\chi}_{+,R}+\tilde{\chi}_{+,L}}{\sqrt{K_+}}&=&0 \nonumber \\
\frac{\tilde{\chi}_{0,R}+\tilde{\chi}_{0,L}}{\sqrt{K_0}}+\frac{\tilde{\chi}_{+,R}+\tilde{\chi}_{+,L}}{\sqrt{2K_+}}&=&0 \nonumber \\
\sqrt{K_0}(\tilde{\chi}_{0,L}-\tilde{\chi}_{0,R})-\frac{\sqrt{K_+}}{\sqrt{2}}(\tilde{\chi}_{+,L}-\tilde{\chi}_{+,R})+\frac{\tilde{\chi}_{-,L}+\tilde{\chi}_{-,R}}{\sqrt{2K_-}} &=&0
\eea
This immediately leads to the following solution
\bea
\tilde{\chi}_{0,R}&=&-\tilde{\chi}_{0,L},\ \ \tilde{\chi}_{+,R}=-\tilde{\chi}_{+,L}\nonumber \\
\tilde{\chi}_{-,R}&=&-2\sqrt{2K_0K_-}\tilde{\chi}_{0,L}+2\sqrt{K_+K_-}\tilde{\chi}_{+,L}-\tilde{\chi}_{-,L}
\eea
Therefore, the rotation matrix is given by
\bea
{\cal O}_{+}^{(g)}=
\begin{pmatrix}
	-1 & 0 & 0 \\
	0 & -1 & 0 \\
	-2\sqrt{2K_0K_-} & 2\sqrt{K_+K_-} & -1 \\
\end{pmatrix}.
\eea
It is easy to verify that ${\cal O}_+^{(g)}$ fails to satisfy the orthogonal condition. As a result, $\chi_+$ fixed point is not physical in the presence of $g$, and should not be considered as a candidate of fixed point in the phase diagram. Similar argument can be performed to other fixed points, and we find that $\chi_{\pm}$ and $A_{1,2}$ fixed points are unphysical. The analysis of remaining physical fixed points are given below.

\subsection{N2. Disconnecting fixed point}
With $n_-=0$, we find that the disconnecting fixed point is characterized by the same ${\cal O}_D$ matrix,
\bea
{\cal O}_D=\begin{pmatrix}
	-1 & 0 & 0 \\
	0 & -1 & 0 \\
	0 & 0 & -1 \\
\end{pmatrix}.
\eea
The explicit boundary condition is
\bea
\phi_0=\phi_+=\phi_-=0,
\eea
where both $\phi_-$ and $\theta_-$ are pinned at $x=0$. Therefore,
\bea
\Delta(T^s_{10})&=&\frac{1}{4}(2K_0+K_+) \nonumber \\
\Delta(T^s_{01})&=&\frac{1}{4}(2K_0+K_+) \nonumber \\
\Delta(T^p_{01})&=&2K_0+K_+ \nonumber \\
\Delta(T^p_{12})&=&0 \nonumber \\
\Delta(T^p_{20})&=&2K_0+K_+ \nonumber \\
\Delta(B_{0})&=&0 \nonumber \\
\Delta(B_{1})&=&0 \nonumber \\
\Delta(B_{2})&=&0.
\eea
Notice that the scaling dimensions have been modified compared with Eq. \ref{Eq: Delta of DFP}. Here $\Delta(T^p_{12})=0$ is a manifestation of pinning $\theta_-$, which will also appear in other fixed points. Now, disconnecting fixed point is stable when
\bea
K_0&>&\frac{4-K_+}{2} \nonumber \\
\label{Eq: DFP stability with theta pinning}
\eea

\subsection{N3. Pair-tunneling fixed point}
With $n_-=0$, the ${\cal O}_D$ matrix remain unchanged compared with the $n_-\neq 0$ case.
Thus, we still have
\bea
\phi_0=-\sqrt{2}\phi_+,\ \theta_+=\sqrt{2}\theta_0, \ \theta_-=0,
\eea
where $\phi_-$ is not pinned by the boundary condition, and thus will be strongly fluctuating due to the pinning of $\theta_-$. Therefore, terms that involve $\phi_-$, such as $T_{\pm}^s$ and $B_{1,2}$, will vanish under RG. The only allowed perturbation term at this fixed point is the backscattering term  $B_0$, with
\bea
\Delta(B_{0})=\frac{4}{2K_0+K_+}
\eea
So, pair-tunneling fixed point is stable when
\bea
K_0<\frac{4-K_+}{2}
\label{Eq: PTFP stability with theta pinning}
\eea
Comparing this stability condition with Eq. \ref{Eq: DFP stability with theta pinning}, we find that disconnecting fixed point and pair-tunneling fixed point are dual to each other.

\subsection{N4. $A_0$ fixed point}
For asymmetric fixed point $A_0$, ${\cal O}_{A_0}$ remains the same with or without the pinning condition $n_-=0$. The explicit boundary condition is
\bea
\phi_0=\phi_+=\theta_-=0.
\eea
Then $\phi_-$ is also strongly fluctuating, and the non-vanishing tunneling processes are pair-tunneling $T^p_{01}$ and $T^p_{20}$, with
\bea
\Delta(T^p_{01})=\Delta(T^p_{20})=2K_0+K_+ \nonumber \\
\eea
Thus, $A_0$ fixed point is stable when
\bea
2K_0+K_+>1
\eea

\subsection{N5. Phase diagram with rMZM}

\begin{figure}[t]
	\centering
	\includegraphics[width=0.6\textwidth]{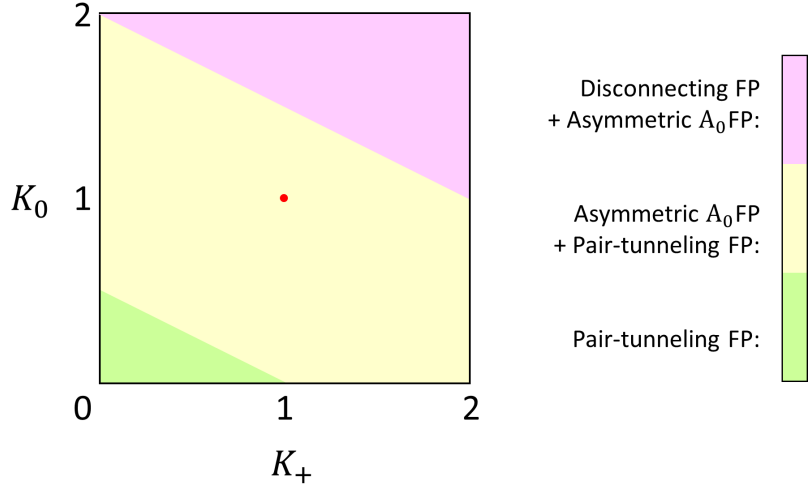}
	\caption{The phase diagram with rMZM (pair hopping interaction).}
	\label{Fig: PD with pair hopping}
\end{figure}

The phase diagram can be easily mapped out in Fig. \ref{Fig: PD with pair hopping}, and quite remarkably, the phase boundaries are completely independent of the value of $K_-$, in comparison with Fig. \ref{Fig: PD without pair hopping}. As shown in Fig. \ref{Fig: PD with pair hopping}, in the weak interacting region, the DSM nanowire junction is in the coexisting phase between $A_0$ fixed point and pair-tunneling fixed point. This signals the emergence of an unstable fixed point which characterizes the phase transition between $A_0$ fixed point and pair-tunneling fixed point, which is similar to a pinch-off transition.

\section{Appendix O. A further look at the ${\cal O}$ matrix}
In the previous section, we have successfully mapped out the tunneling phase diagrams of the 2LL/LL junction with and without the pair-hopping interaction. The key information of each tunneling phase is encoded in the orthogonal matrix ${\cal O}$. Interestingly, the form of ${\cal O}$ of three-wire-junctions has been discussed in earlier literatures \cite{bellazzini2007,bellazzini2008,bellazzini2009,agarwal2009} under very general assumptions. In particular, it has been shown that ${\cal O}$ matrices could be classified into two different classes based on whether their determinant is $+1$ or $-1$:
\begin{itemize}
	\item When det ${\cal O}=-1$, ${\cal O}$ corresponds to a fixed point that possesses a $Z_3$ symmetry among the three quantum wires. For example, at the disconnecting fixed point (DFP), every quantum wire is in the strong backscattering limit. If we relabel the wire indices at DFP, the DFP physics will not be affected and thus possesses the $Z_3$ symmetry among three quantum wires. Other fixed points that belong to this class are: pair-tunneling fixed point and chiral $\chi_{\pm}$ fixed points.
	\item When det ${\cal O}=+1$, ${\cal O}$ corresponds to a fixed point that explicitly breaks the $Z_3$ symmetry among the three quantum wires. The asymmetric fixed points $A_{0,1,2}$ belong to this class. If we start from $A_0$FP and relabel the wire indices in the following way: $0\rightarrow 1,\ 1\rightarrow2,\ 2\rightarrow0$, we will arrive at the $A_1$FP. Therefore, $A_0$FP, as well as $A_1$ and $A_2$FP, explicitly breaks the $Z_3$ symmetry.
\end{itemize}
To compare with the results in Ref. \cite{bellazzini2009,agarwal2009}, we first notice that there are three important differences between the notations:
\begin{enumerate}
	\item In Ref. \cite{bellazzini2009}, the three quantum wires share the same Luttinger parameter $g$. While in our model, we have assumed the Luttinger parameters $K_0$ and $K_{\pm}$ to be different from each other.
	\item In our theory, we have performed a unitary transformation to the dual boson fields, and rotate $\phi_{1,2}$ and $\theta_{1,2}$ into the bonding/anto-bonding basis $\phi_{\pm}$ and $\theta_{\pm}$:
	\bea
	U=
	\begin{pmatrix}
		1 & 0 & 0 \\
		0 & \frac{1}{\sqrt{2}} & \frac{1}{\sqrt{2}} \\
		0 & \frac{1}{\sqrt{2}} & -\frac{1}{\sqrt{2}} \\
	\end{pmatrix}
	\eea
	\item The bosonization convention in our work is different from that of earlier works:
	\bea
	\psi_{l,R}\sim e^{i\phi_{l,O}}\sim e^{i\sqrt{\pi}\chi_{l,R}},\ \ \psi_{l,L}\sim e^{i\phi_{l,I}}\sim e^{-i\sqrt{\pi}\chi_{l,L}}
	\eea
	where $\phi_{l,O/I}$ and $\chi_{l,R/L}$ are the chiral bosons defined in Ref. \cite{bellazzini2009} and in our work, respectively. As a result, we have
	\bea
	\phi_{l,O}=\sqrt{\pi}\chi_{l,R},\ \ \phi_{l,I}=-\sqrt{\pi}\chi_{l,L}.
	\label{Eq: chiral field of S matrix}
	\eea
\end{enumerate}

\begin{table}[t]
	\begin{tabular}{cccccccc}
		\toprule[1.5pt] \\
		& DFP & $\chi_+$FP & $\chi_-$FP & PFP & A$_0$FP & A$_1$FP & A$_2$FP \\[0.5em]
		\midrule \hline \\
		${\cal O}\ $: & ${\cal O}_D$ & ${\cal O}_+$ & ${\cal O}_-$  & ${\cal O}_P$  & ${\cal O}_{A_0}$ & ${\cal O}_{A_1}$ & ${\cal O}_{A_2}$ \\
		\\[-0.5em] \hline \\[-0.5em]
		${\cal S}\ $: & ${\cal S}^{(1)}(t=0)\ $ & ${\cal S}^{(1)}(t=-\frac{2\pi}{3})\ $ & ${\cal S}^{(1)}(t=\frac{2\pi}{3})\ $  & ${\cal S}^{(1)}(t=-\pi)\ $ & ${\cal S}^{(2)}(t=\pi)\ $ & ${\cal S}^{(2)}(t=\frac{\pi}{3})\ $ & ${\cal S}^{(2)}(t=-\frac{\pi}{3})$ \\
		\\
		\bottomrule[1.5pt]
	\end{tabular}
	\caption{In this table, we applied Eq. \ref{Eq: O matrix and S matrix} and established the mapping between ${\cal O}$ matrices we found using DEBC method in our 2LL/LL junction and the ${\cal S}$ matrices defined in Ref. \cite{bellazzini2009}.}
	\label{Table:O and S matrix}
\end{table}

As a comparison, we follow the convention of Ref. \cite{bellazzini2009} and denote the orthogonal matrix that connects $\phi_O$ and $\phi_I$ as ${\cal S}$. Ref. \cite{bellazzini2009} showed the explicit expressions of ${\cal S}$ matrix parametrized by a single angular variable $t\in [-\pi,\pi)$ at $g=1$:
\bea
{\cal S}^{(1)}(t)&=&\frac{1}{3}
\begin{pmatrix}
	1+2\cos t & 1-\cos t +\sqrt{3}\sin t & 1-\cos t -\sqrt{3}\sin t \\
	1-\cos t -\sqrt{3}\sin t & 1+2\cos t & 1-\cos t +\sqrt{3}\sin t \\
	1-\cos t +\sqrt{3}\sin t & 1-\cos t -\sqrt{3}\sin t & 1+2\cos t \\
\end{pmatrix} \nonumber \\
{\cal S}^{(2)}(t)&=&\frac{1}{3}
\begin{pmatrix}
	1-2\cos t & 1+\cos t -\sqrt{3}\sin t & 1+\cos t +\sqrt{3}\sin t \\
	1+\cos t -\sqrt{3}\sin t & 1+\cos t +\sqrt{3}\sin t  & 1-2\cos t \\
	1+\cos t +\sqrt{3}\sin t  & 1-2\cos t & 1+\cos t -\sqrt{3}\sin t \\
\end{pmatrix}
\eea
where ${\cal S}^{(1)}$ with det ${\cal S}^{(1)}=+1$ and ${\cal S}^{(2)}$ with det ${\cal S}^{(2)}=-1$ denote the $Z_3$-symmetric and $Z_3$-asymmetric classes, respectively. Based on our previous analysis, by taking $K_0=K_+=K_-=1$ and performing a unitary transformation, we are able to map our ${\cal O}$ matrices found in the previous sections to ${\cal S}^{(1)}$ and ${\cal S}^{(2)}$:
\bea
-(U{\cal O}_\alpha U^{-1})|_{K_0=K_{\pm}=1}={\cal S}^{(\beta)}(t)
\label{Eq: O matrix and S matrix}
\eea
where $\alpha$ is the subscript for ${\cal O}$ matrix depending on which fixed point we are talking about. $\beta=1,2$ is the superscript for ${\cal S}$ matrix. The minus sign in Eq. \ref{Eq: O matrix and S matrix} comes from the definition of $\phi_{l,I}$ in Eq. \ref{Eq: chiral field of S matrix}. Here we would like to point out that ${\cal O}_{\alpha}$ and ${\cal O}_{\alpha}|_{K_0=K_{\pm}=1}$ only differ by a field rescaling process in the definition. Therefore, it is sufficient to discuss the relationship between the non-interacting limit of ${\cal O}_{\alpha}$ and ${\cal S}^{(\beta)}$, which contains all of the key information. With Eq. \ref{Eq: O matrix and S matrix}, we have mapped all ${\cal O}$ matrices to the corresponding ${\cal S}$ matrices, and summarized these results in Table. \ref{Table:O and S matrix}.


\end{document}